\documentclass[twocolumn, superscriptaddress,aps,pre, 10pt]{revtex4-2}

\usepackage[utf8]{inputenc}
\usepackage{amsfonts}
\usepackage{amsmath}
\usepackage{amssymb}

\usepackage{ulem}
\usepackage[super]{nth}
\usepackage{tabularx}    
\usepackage{multirow}    
\usepackage{longtable}   

\usepackage{booktabs}    
\usepackage[export]{adjustbox}
\usepackage{url}
\usepackage[colorlinks=true, allcolors=blue]{hyperref}
\usepackage{enumitem}

\usepackage{graphicx}
\usepackage{orcidlink}

\begin{document}

\title{Atomistic study of dislocation formation during Ge epitaxy on Si}

\author{Luis Martín-Encinar \orcidlink{0000-0002-4839-7604}}\email[Corresponding author: ]{luis.martin.encinar@uva.es}\affiliation{Dpto. of Electricity and Electronics, E.T.S.I. de Telecomunicación, Universidad de Valladolid,47011,Valladolid,Spain}

\author{Luis A. Marqués \orcidlink{0000-0002-9269-4331}} \affiliation{Dpto. of Electricity and Electronics, E.T.S.I. de Telecomunicación, Universidad de Valladolid,47011,Valladolid,Spain}

\author{Iván Santos \orcidlink{0000-0003-1388-4346}} \affiliation{Dpto. of Electricity and Electronics, E.T.S.I. de Telecomunicación, Universidad de Valladolid,47011,Valladolid,Spain}

\author{Lourdes Pelaz \orcidlink{0000-0001-7181-1079}} \affiliation{Dpto. of Electricity and Electronics, E.T.S.I. de Telecomunicación, Universidad de Valladolid,47011,Valladolid,Spain}\makeatletter
\def\maketitle{
\@author@finish
\title@column\titleblock@produce
\suppressfloats[t]}
\makeatother

\date{\today}

\begin{abstract}  
We performed classical molecular dynamics simulations to investigate, from an atomistic point of view, the formation of dislocations during the epitaxial growth of Ge on Si. We show that simulations at 900 and 1000 K with deposition rates of 10$^8$ monolayers per second provide a good compromise between computational cost and accuracy. In these conditions, the ratio between the Ge deposition rate and the ad-atom jump rate is analogous to that of out-of-equilibrium experiments. In addition, the main features of the grown film (intermixing, critical film thickness, dislocation typology, and surface morphology) are well described. Our simulations reveal that dislocations originate in low-density amorphous regions that form under valleys of the rough Ge film surface. Atoms are squeezed out of these regions to the surface, releasing the stress accumulated in the film and smoothing its surface. Amorphous regions grow until atoms begin to rearrange in dislocation half-loops that propagate throughout the Ge film. The threading arm ends of the dislocation half-loops move along the surface following valleys and avoiding islands. The film surface morphology affects the propagation path of the dislocation half-loops and the resulting dislocation network.
\end{abstract}


\maketitle

\section{Introduction}

The excellent properties of Silicon-Germanium (SiGe) and its compatibility with silicon-based technologies have promoted its use in different areas such as photonic, microelectronics, or quantum devices~\cite{Ryzhak2024}. The heteroepitaxial growth of high-Ge-content SiGe or pure Ge layers on Si substrates entails compressive stress in the grown film due to the lattice mismatch between Si and Ge. Depending on growth conditions, the accumulated strain is relaxed elastically by the formation of 3D islands after the formation of a wetting layer~\cite{ARAPKINA2023}, or plastically through the generation of misfit dislocations (MDs), threading dislocations (TDs) and partial dislocations (PDs) within the film~\cite{Du2021}.

Many experimental works have been devoted to the study of dislocations in Ge/Si systems, as they degrade the structural and electronic properties of the material. The main type of observed dislocations are 60$^{\circ}$ and 90$^{\circ}$ MDs, being the latter the most effective one for releasing strain~\cite{BOLKHOVITYANOV2015}. It has been proposed that 90$^{\circ}$ MDs generate through the coalescence of complementary pairs of 60$^{\circ}$ MDs~\cite{BOLKHOVITYANOV2012_1}, although some authors suggested its direct formation on the strained Ge film~\cite{WIETLER2008}. Dislocations originate as half-loops in the growing film~\cite{bolkhovityanov2004}, and their nucleation has been related to stress concentrators associated to cups and steps on the surface~\cite{Jesson1993, tersoff1994}. Point defects might also play a role in dislocation formation~\cite{Pichaud2009}, as they do in dislocation motion~\cite{Barbisan2022}. Generated dislocations can also be the source of new dislocations through multiplication mechanisms~\cite{Capano92}. However, there is still a lack of understanding about the initial stages of dislocation nucleation at the atomic level, as experimental techniques do not allow direct observation of such processes with sufficient spatial and temporal resolution. In this context, the use of computational methods can be helpful.

Among the available computational techniques, classical molecular dynamics (CMD) allows the atomistic dynamic simulation of large enough systems to accommodate dislocations. However, due to time scale constraints, the typical experimental deposition rates cannot be directly simulated using CMD. Consequently, most CMD works on heteroepitaxial growth in the literature are either limited to very simple system cases (2D systems and elementary interatomic potentials~\cite{dong1998}), or to simulations where dislocations are introduced \textit{ad hoc}~\cite{Barbisan2022,maras,Maras2017,Trushin_2016}), or their nucleation is forced by incorporating surface steps~\cite{zhang2018,izumi2010,Li2012}. While these kinds of simulations offered significant information on dislocation propagation, interaction among them or with point defects, they did not provide detailed insights into the early stages of dislocation formation.

In this work, we use CMD to elucidate the atomistic origin of dislocations through the direct simulation of the heteroepitaxial growth process of Ge on Si(001). We previously analyze the simulation conditions that can provide information comparable to that of experiments, and validate our CMD simulations by comparing different characteristics of the grown film with experimental observations.

\section{Simulation details}\label{sec::methods}

To perform our CMD study, we used the Large-scale Atomic/Molecular Massively Parallel Simulator (LAMMPS)~\cite{lammps}. We employed the Open Visualization Tool (OVITO)~\cite{Stukowski_2010} to visualize and analyze atomic simulation data, in particular, to identify crystallographic structures and dislocations. We use throughout the paper the OVITO terminology (see Appendix~\ref{appendixA}).

Atomic interactions were described by the Stillinger-Weber (SW) potential with the parametrizations of Balamane \textit{et al.}~\cite{balamane} for Si and Posselt \textit{et al.}~\cite{swp} for Ge, with Gilmer and Grabow mixing rules~\cite{gilmergrabow}, as this particular combination is the best for describing surface dynamics and intermixing on the Si-Ge system~\cite{luis2023}.

In our simulation cell, the Si substrate consisted of a perfectly crystalline Zinc-Blende lattice containing 27648 atoms, with a thickness of 12 monolayers (MLs) along the $Z$ direction. Substrate dimensions were 24$\sqrt{2} a_T^{Si}$ $\times$ 24$\sqrt{2} a_T^{Si}$ $\times$ 3 $a_T^{Si}$, being $a_{T}^{Si}$ the Si equilibrium lattice constant at temperature $T$ and zero applied stress. Cell axes $X$, $Y$ and $Z$ were oriented along the [110], [-110], and [001] directions, respectively. Atoms of the topmost layer were arranged along $Y$ to form the typical 2$\times$1 Si(001) surface reconstruction. Periodic boundary conditions were introduced along $X$ and $Y$ directions, while free boundary conditions were applied in the $Z$ direction. 

The Si substrate was divided into three zones: fixed layers, thermal bath and surface region. The four atomic layers at the bottom were fixed to mimic bulk behavior and prevent the system from moving as a whole. Atoms in the middle four layers were used to create a thermal bath to control the system temperature. Atoms in the surface region and thermal bath were free to move according to Newton's dynamics. Initially, atoms in the thermal bath were given random velocities in a Maxwell-Boltzmann distribution corresponding to the desired simulation temperature, and then the whole system was equilibrated for 10$^5$ steps before initiating the deposition of Ge. Integration timestep was set to 2 fs.

Ge atoms were deposited from a height of 85$\textup{~\AA}$ above the substrate surface on a random ($X$, $Y$) position. Each Ge atom had an initial velocity of -0.7 nm/ps in the $Z$ direction, corresponding to an energy of 0.185 $e$V (low enough to prevent substrate damage or interdiffusion into the Si substrate). We simulated the deposition of a total of 40 Ge MLs (92160 atoms) at temperatures of 650, 800, 900 and 1000 K. To keep the substrate at the desired temperature, atom velocities were scaled within the thermostat layers every 2000 steps. Once all Ge atoms were deposited, the system sample was equilibrated for 10$^5$ additional steps. To make a statistical study, we performed five different simulations for each temperature by changing the random number seed used to select the surface impact point of the Ge atom and to generate the initial Maxwell-Boltzmann velocity distribution.

Typical deposition rates in real epitaxial growth processes range between 0.001 and 10 MLs/s~\cite{CARLSSON2010314,FRIGERI2011480}, depending on the desired properties of the film. However, CMD can only reach simulation times on the order of microseconds at most, which implies that the slowest deposition rate affordable in the simulations is around $10^6$ MLs/s. With this premise, CMD seems unable to simulate Ge epitaxial growth on Si in experimental conditions.

The two main factors that control film growth are the Ge deposition rate $r_{dep}$ and the atomic surface diffusivity. When $r_{dep}$ is low, each deposited Ge ad-atom can explore wide areas on the substrate surface and crystallize in the most favorable epitaxial positions (minimum-energy positions matching the underlying crystal lattice) before the arrival of the next ad-atom. If $r_{dep}$ increases, the time between successive deposited Ge atoms decreases, ad-atoms move short distances on the Si surface, and tend to crystallize in kinetically trapped structures. This limitation can be compensated by increasing the temperature, which exponentially enhances diffusivity, and thus the Ge ad-atom can explore more epitaxial positions before the arrival of the subsequent Ge atom. Consequently, to compare experimental and CMD conditions, the relevant figure is the ratio between the deposition rate $r_{dep}$ and the jump rate of ad-atoms, $\nu_d(T)$.

\begin{figure}[htb!]
	\centering
	\includegraphics[width=\columnwidth]{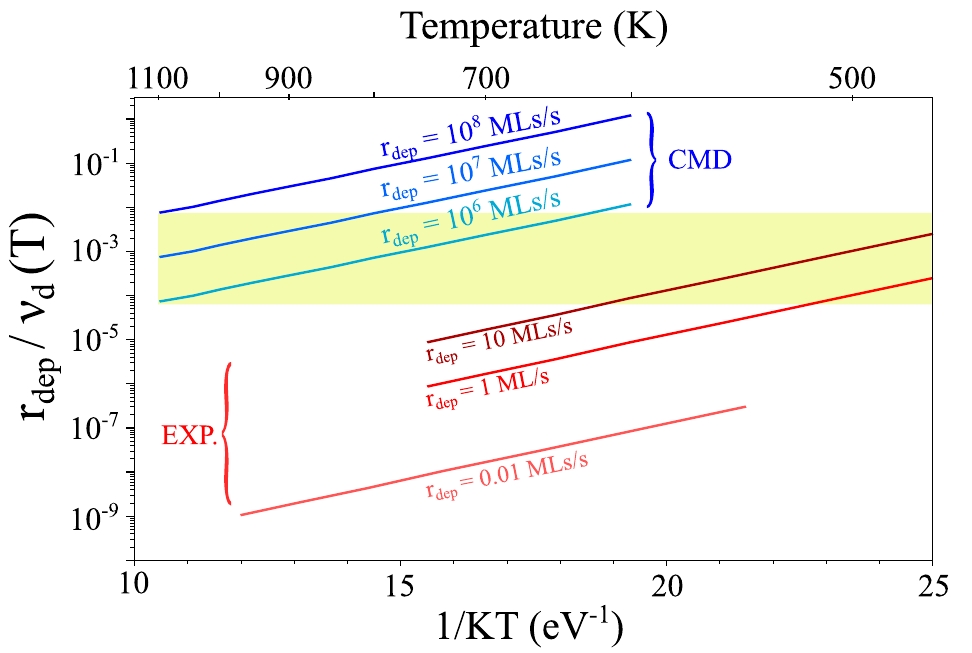}
    \caption{Comparative of the ratio between deposition rate \texorpdfstring{$r_{dep}$}{r_dep} and jump rate of ad-atoms \texorpdfstring{$\nu_d(T)$}{nu_d(T)} for typical conditions of CMD simulations (blue lines) and experiments (red lines). The yellow band represents conditions where CMD simulations and experimental conditions are comparable.}
	\label{fig:figure1}
\end{figure}

In Figure~\ref{fig:figure1}, we plot the ratio between $r_{dep}$ and $\nu_d(T)$ for typical CMD and experimental growth conditions of Ge on Si. For CMD, we consider $\nu_d(T)$ calculated in Ref.~\cite{luis2023}, and deposition rates of 10$^6$, 10$^7$ and 10$^8$ MLs/s. For experimental conditions, we take values for $\nu_d(T)$ extracted from Ref.~\cite{lagally}, and consider deposition rates of 0.01, 1 and 10 MLs/s. Typically, equilibrium conditions involve low deposition fluxes (e.g. 0.001-0.1 MLs/s) and high temperatures ($\sim$900 K), leading to the formation of 3D islands \cite{RADIC2006,Capellini2003}. In contrast, out-of-equilibrium conditions are characterized by high deposition fluxes (between 0.1 and 10 MLs/s) and relatively low temperatures (around 400-500 K), leading to the formation of smooth Ge films with dislocations \cite{CHOI2008,SHAH2011}.

According to Fig.~\ref{fig:figure1}, experimental conditions typical of 3D islanding are not achievable with CMD simulations. Nevertheless, with CMD at $r_{dep}$ of 10$^6$ MLs/s at temperatures between 700 and 1100 K, or at 10$^7$ MLs/s between 900 and 1100 K, the conditions are comparable to those of experiments that lead to the formation of smooth Ge films with dislocations around 500-600 K. The deposition rate of 10$^8$ MLs/s is at the limit of the CMD simulation conditions comparable to those of experiments, but it is computationally more affordable for large systems. 

To check the validity of the use of such a high $r_{dep}$ value, we carried out two test cases. First, we made sure that even the fastest $r_{dep}$ did not induce unrealistic high-energy configurations, pronounced islands or stacking faults in a non-strained Ge homosystem. For that, we simulated the deposition of 54 Ge MLs on Ge(001) 2$\times$1 at $r_{dep}$ values of 10$^6$, 10$^7$ and 10$^8$ MLs/s, and at a temperature of 1000 K. In all simulations, we obtained Ge layers coherently grown on the Ge substrate according to the Frank–van der Merwe growth mode. Second, we simulated the deposition of Ge on Si(001) 2$\times$1 using $r_{dep}$ of 10$^8$ MLs/s and compared the results with those at a lower $r_{dep}$ of 10$^7$ MLs/s, at 900 and 1000 K. We did not find significant differences in dislocation formation, their density and origin, but calculation times are ten times larger for the slower rate. Consequently, we will use $r_{dep}$=10$^8$ MLs/s in our study, as it offers a good compromise between accuracy and computational cost. In any case, the validity of CMD simulations will be determined by comparing with experiments whenever possible.

\section{Characterization of the Ge grown film}\label{sec::results}

During the initial stages of Ge deposition on Si(001), we found that some Ge atoms randomly replaced Si atoms which were incorporated within the first deposited Ge layers, leading to Ge-Si intermixing. In conditions where no dislocations had been formed yet, the Ge content in the original surface layer of the Si substrate ranged from 13 to 25\%, in accordance with experimental observations~\cite{ikeda}, DFT calculations~\cite{cho}, and kMC simulations~\cite{wagner}, which give Ge-Si intermixing values from 10 to 20\%. We noticed that intermixing increased with temperature because atomic mobility is enhanced and the ratio between surface diffusion and intermixing frequencies decreases~\cite{luis2023}. When dislocations nucleated and propagated down to the Si/Ge interface, severe atomic rearrangement was produced. In these cases, we found 35\% Ge atoms in the original surface layer of the Si substrate, and $\sim$10\% and $\sim$5\% in the subsequent two layers below. This observation agrees with experiments, where a higher Ge content was found in layers below the Si surface when formed dislocations reach the Si/Ge interface~\cite{ARROYO2019}. For further details, see Fig.~\ref{fig:ch5:coverage} in Supplementary Material.

The initially grown Ge layers were almost flat and ``coherent" with the underlying Si substrate, until a critical Ge film thickness was reached, and defective regions and dislocations were formed. For each temperature, the critical film thickness varied widely from one simulation to another, with a tendency to decrease as temperature increased (see Fig.~\ref{fig:ch5:criticalthickness} in Supplementary Material). In our CMD simulations at 1000 K, we obtained critical thickness values between 10 and 20$\textup{~\AA}$, in agreement with theoretical models~\cite{mattews1975,People1985} and experiments~\cite{bean1984,houghton1990m}.

\begin{figure*}[htb!]
    \centering
    \includegraphics[width=\textwidth]{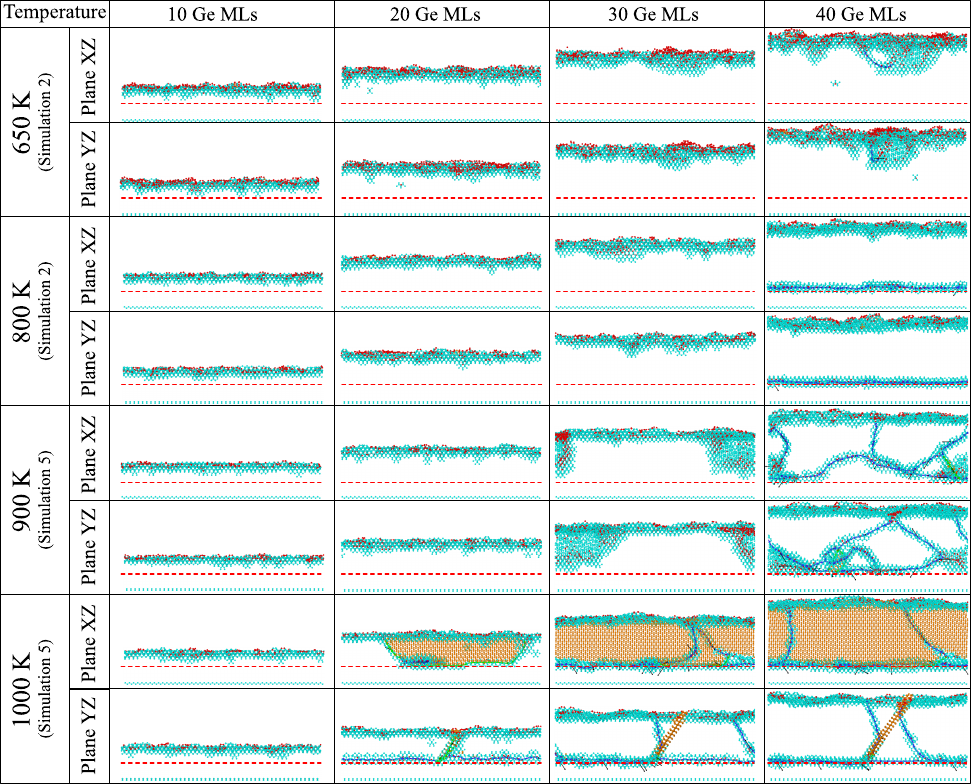}
    \caption{Side views of samples during the Ge epitaxial growth process at different temperatures. The corresponding simulated case of Table~\ref{tab:table1} is indicated in parenthesis. Only non-cubic diamond atoms are shown. Atoms are colored according to their local structure: cubic diamond up to \nth{1} or \nth{2} neighbors (\protect\includegraphics[height=1.5ex]{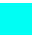}), hexagonal diamond (\protect\includegraphics[height=1.5ex]{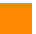}), and \textit{other} (\protect\includegraphics[height=1.5ex]{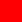}). Dark blue and green lines indicate dislocations with (a/2) $<$110$>$ and (a/6) $<$112$>$ Burgers vectors (black arrows), respectively. Horizontal red dashed lines indicate the original position of the Si surface substrate.}
    \label{fig:figure2}
\end{figure*}

Table~\ref{tab:table1} collects the variety of dislocations formed after depositing 40 Ge MLs, illustrating the stochastic nature of dislocation nucleation. Figure~\ref{fig:figure2} shows side snapshots taken during growth for representative simulations at each temperature. Although there is a variety of situations, we recognize some general trends. At low temperatures, dislocations were in an incipient state or even not formed at all, due to the slow dynamics. At higher temperatures, denser dislocation networks formed, composed of either TDs, 60$^{\circ}$ MDs or 90$^{\circ}$ MDs, in agreement with experimental findings~\cite{ARROYO2019}. This behavior is consistent with dislocation formation being a thermally activated process~\cite{janzen2001}. We also observed the formation of PDs, characterized by the presence of Stacking Faults (SFs) that adopted a hexagonal diamond structure and tended to cover larger areas as temperature increased. In some cases, the threading arms of single MDs were annihilated due to boundary conditions. The reduction of the stress in the deposited Ge film is related to the amount, type and arrangement of the formed dislocation network (see section S3 in Supplementary Material).

In our CMD simulations, we noticed several ways to generate 90$^{\circ}$ MDs. In most cases, 90$^{\circ}$ MDs formed directly without the prior presence of 60$^{\circ}$ MDs. Sometimes (e.g. simulations 4 or 5 at 900 K), 60$^{\circ}$ MDs are combined with 90$^{\circ}$ MDs. In other cases, more frequently at higher temperatures, 90$^{\circ}$ MDs are associated to the generation of Shockley PDs. These formation mechanisms resemble those proposed by other authors~\cite{BOLKHOVITYANOV2012_1, WIETLER2008, maras}, although they are not so clean and ideal, as they show great complexity and variability among different simulation runs.

\begin{figure}[t!]
    \centering
    \includegraphics[width=\columnwidth]{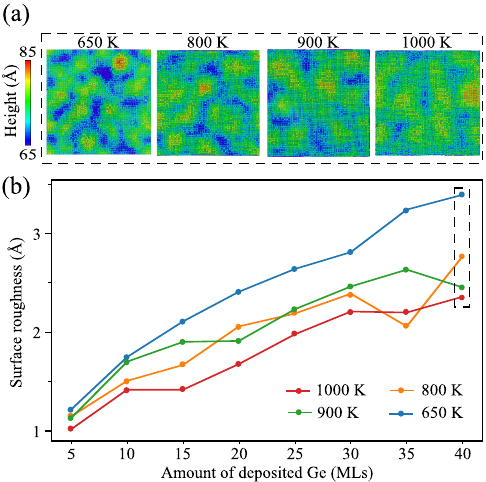}
    \caption{(a) Top views of the samples corresponding to the simulations in Fig.~\ref{fig:figure2} after depositing 40 Ge MLs. Atoms are colored according to their height. (b) Surface roughness as a function of the number of deposited Ge MLs at different temperatures for the simulation cases of Fig.~\ref{fig:figure2}.}
    \label{fig:roughness}
\end{figure}

Surface morphology was affected by temperature and dislocation formation as Ge deposition proceeded. The roughness of the surface film was computed as follows:

\begin{equation}
Roughness = \sqrt{\frac{\sum_{i=0}^{n} \left( Z_{i} - \overline{Z}  \right)^{2}}{n}}\label{eq:roughness}       
\end{equation}

\noindent
where \textit{Z\textsubscript{i}} is the Z height of the \textit{i\textsuperscript{th}} surface atom, $\overline{Z}$ is the average surface height, and \textit{n} is the total number of surface atoms. Figure ~\ref{fig:roughness} shows, for the simulations of Fig.~\ref{fig:figure2}, the evolution of roughness as deposition proceeded at different temperatures and top views of the samples after depositing 40 Ge MLs. Larger and higher 3D islands formed on the surface at lower temperatures, while surfaces were smoother for higher temperatures. This is consistent with the enhanced surface diffusivity at higher temperatures, which favors the movement of atoms on the surface to fill denuded zones, and with the increased formation of dislocations which relax the film strain. The influence of dislocations on surface roughness is clearly evidenced, for example, in the simulation at 900 K. When a dislocation formed (after 20 Ge MLs deposited), the surface roughness decreased. As deposition proceeded, the surface roughness increased and it dropped again when a new dislocation nucleated (after 35 Ge MLs deposited). This cyclic behavior is consistent with experimental observations ~\cite{LeGoues1994,RovarisPRB2016}.

\begin{table*}[htb!]
\centering
\caption{Summary of the formed dislocations once deposition was completed at 650, 800, 900 and 1000 K. For simplicity, we denote dislocations using abbreviations: TD (threading dislocation), MD (misfit dislocation), FPD (Frank partial dislocation), SPD (Shockley partial dislocation), and ND (no dislocations or extended defects). The symbol $\perp$ denotes orthogonal dislocations. In parenthesis, we indicate the number of formed dislocations, when there are more than one.}
\label{tab:table1}
\begin{tabular}{c|c|c|c|c|}
\cline{2-5}
                                              & 650 K  & 800 K & 900 K & 1000 K    \\ \hline
\multicolumn{1}{|c|}{\multirow{2}{*}{Simulation 1}} & \multirow{2}{*}{dislocation half-loop}                      & dislocation half-loop                      & two $\perp$ SPDs                     & \multirow{2}{*}{two $\perp$ 90$^{\circ}$ MDs}                      \\
\multicolumn{1}{|c|}{}                              &                                        &   SPD (x2)                                      &  dislocation half-loop                                     &                                         \\ \hline
\multicolumn{1}{|c|}{\multirow{2}{*}{Simulation 2}} & \multirow{2}{*}{dislocation half-loop}                      & \multirow{2}{*}{60$^{\circ}$ MD}                      & \multirow{2}{*}{90$^{\circ}$ MD}                      & \multirow{2}{*}{90$^{\circ}$ MD}                      \\
\multicolumn{1}{|c|}{}                              &     \multicolumn{1}{|c|}{}                                   &                                        &                                        &                                        \\ \hline
\multicolumn{1}{|c|}{\multirow{2}{*}{Simulation 3}} & \multirow{2}{*}{dislocation half-loop}   
& \multirow{2}{*}{dislocation half-loops (x2)}                       & complex dislocation network                     & \multirow{2}{*}{two $\perp$ 90$^{\circ}$ MDs}                      \\
\multicolumn{1}{|c|}{}                              &                                    &       &   (60$^{\circ}$ and 90$^{\circ}$ MDs) ; FPD ; SPD                                     &                                        \\ \hline
\multicolumn{1}{|c|}{\multirow{2}{*}{Simulation 4}} & \multirow{2}{*}{SPDs (x2)} & FPD &  complex dislocation network & SPDs (x2) \\
\multicolumn{1}{|c|}{}                    &          &    dislocation half-loop        &     (60$^{\circ}$ and 90$^{\circ}$ MDs)                &     (Stair Rod dislocation)             \\ \hline
\multicolumn{1}{|c|}{\multirow{2}{*}{Simulation 5}} & \multirow{2}{*}{ND} & \multirow{2}{*}{ND} & complex dislocation network & 60$^{\circ}$ MD with TD \\
\multicolumn{1}{|c|}{}                              &                   &           &  (60$^{\circ}$ and 90$^{\circ}$ MDs)                 & 90$^{\circ}$ MD; SPD                  \\ \hline
\end{tabular}
\end{table*}

\clearpage

We have shown that CMD simulations of Ge deposition on Si, with $r_{dep}$=10$^8$ MLs/s at the high-temperature range (900-1000 K), are able to reproduce the dislocation typology and the main features of the grown film (intermixing, critical thickness, and surface roughness), in agreement with experiments in out-of-equilibrium growth conditions.

\section{Atomistic insight into the origin of dislocations} \label{sec::atdet}

From the inspection of our CMD simulations, we perceived the significant role of surface roughness in the nucleation of dislocations, as highlighted by Tersoff and LeGoues~\cite{tersoff1994}. We found that dislocations nucleated in surface zones among islands, commonly known as valleys or depressions which act as stress concentrators. Nevertheless, the specific depression site on the surface where dislocations nucleated was, in principle, unpredictable. In all our simulations, the region under the depression where dislocations originate is characterized by local structural disorder, as OVITO identified most of their atoms as \textit{other} (see Appendix~\ref{appendixA}). Consequently, we will refer to these zones that preceded dislocations as \textit{disordered regions}.

Considering the roughness and the great amount of micro-islands and valleys on the surface, there was a high density of sites where dislocations could initially nucleate. In fact, we noticed that small disordered regions often appeared simultaneously in different valleys. During deposition, one of these regions, usually the one located in the most extensive valley, evolved to a deeper and larger disordered region with the shape of a half-loop along a specific crystallographic plane (mostly (100), (001) or $\lbrace011\rbrace$). During the growth of the disordered region, we found that some of the Ge atoms inside moved towards the surface, thus generating ``vacancies'' within that region. This contributed to strain relaxation and prevented the evolution of other incipient disordered regions on the surface.

In all our simulations, we found that valleys in the rough surface Ge film played also a significant role in the propagation of dislocations. The ends of the threading arms of the dislocation half-loops, which were also formed by disordered atoms, moved along the surface through the valleys avoiding islands, thus facilitating the ejection of more Ge atoms from the film. The presence of islands in the propagation pathway of dislocations forced them to change their expansion direction or even to split, leading to dislocation multiplication. Thus, specific surface morphology affects the resulting dislocation network (see supplementary material section S4).

\begin{figure*}[htb!]
    \centering
    \includegraphics[width=\textwidth]{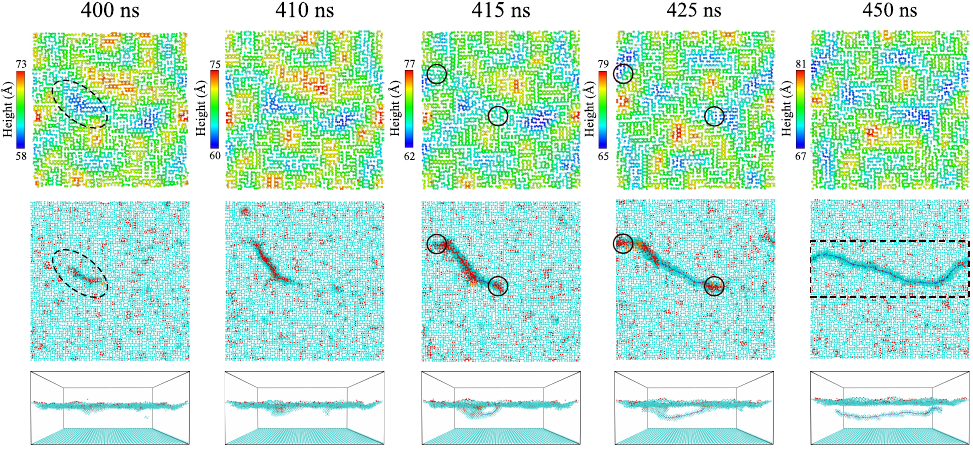}
    \caption{Sample snapshots taken at different times during one of the deposition simulations carried at 900 K (simulation 2 of Table~\ref{tab:table1}). Upper panels are top views where atoms are colored according to their height. Middle panels are also top views where atoms are colored according to their local structure: cubic diamond up to \nth{1} or \nth{2} neighbors (\protect\includegraphics[height=1.5ex]{lightblue.png}), hexagonal diamond (\protect\includegraphics[height=1.5ex]{orange.png}), and \textit{other} (\protect\includegraphics[height=1.5ex]{red.png}) (perfect cubic diamond atoms are not shown). Bottom panels are perspective views of atoms shown in middle panels. Dashed ovals indicate the valley and disordered region where the dislocation starts to form. Full circles highlight the position of dislocation threading arm ends. Dashed rectangle indicates the top surface of the parallelepiped region where the number of disordered atoms and vacancies are computed (shown in fig.~\ref{fig:disordervac}). Dark blue lines are dislocation lines as identified by OVITO.}
    \label{fig:nucleation}
\end{figure*}

In order to illustrate the process of dislocation formation, in Fig.~\ref{fig:nucleation} we show several snapshots taken during one of the Ge deposition simulations carried out at 900 K (simulation 2 of Table~\ref{tab:table1}). At t=400 ns, $\approx$ 35 MLs deposited, islands (red regions in upper panels) with a height of around 1.5 nm with respect to valleys (dark blue regions in upper panels) formed, and an incipient disordered region (red atoms in middle and bottom panels) generated in one of the valleys. At t=410 ns, $\approx$ 35.5 MLs deposited, the disordered region increased its size along the (100) crystallographic plane, aligned with the valley in which it laid in. At t=415 ns, $\approx$ 36 MLs deposited, the disordered zone rearranged adopting the shape of a half-loop lying on a $\lbrace111\rbrace$ plane. Later, the threading arms of the half-loop moved along the surface through the valleys, avoiding islands, which produced kinks in the generated 90 MD line. Eventually, both dislocation threading arms met due to periodic boundary conditions at t=450 ns, $\approx$ 39 MLs deposited. Under additional annealing, this line defect transformed into a perfectly straight 90$^{\circ}$ MD (see supplementary material section S5).

\begin{figure}[htb!]
    \centering
    \includegraphics[width=\columnwidth]{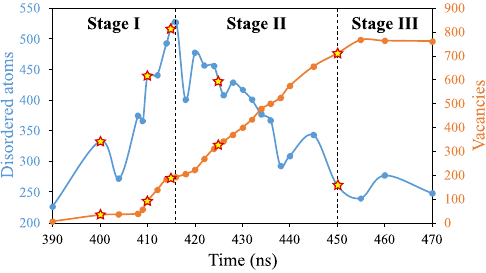}
    \caption{Evolution of the number of \textit{other} atoms and vacancies in a parallelepiped region encompassing the dislocation formed after deposition at 900 K (simulation 2 of Table~\ref{tab:table1}). Stars indicate the times corresponding to the snapshots shown in ~\ref{fig:nucleation}. See text for details.}
    \label{fig:disordervac}
\end{figure}

Figure~\ref{fig:disordervac} shows the evolution of the number of disordered atoms (identified as \textit{other} atoms in OVITO) and ``vacancies'' (missing atoms with respect to the perfect lattice), which have been accounted for within a parallelepiped region encompassing the dislocation in the previously described simulation (dashed rectangle in Fig.~\ref{fig:nucleation} indicates its top surface, its volume goes from the film surface down to the Si/Ge interface). Three different stages are distinguished:
\begin{itemize}
\item 
Stage I corresponds to the generation and growth of the initial disordered region below the surface valley of the Ge film. The initial number of disordered atoms correlates to the amount of \textit{other} atoms which are always present on the film surface. Along with the increase in the number of disordered atoms, there is also a sharp increase in the number of vacancies within the region, which is a natural mechanism that helps to accommodate mismatch and release strain energy~\cite{bauer2006,bulatov,MYRONOV201837}. These missing atoms were squeezed out from the disordered region, moved upward and refilled the valley, thus smoothing the surface.
\item
Stage II corresponds to the formation and propagation of the dislocation half-loop. Atoms in the core of the disordered region rearranged according to lattice sites, which produced a reduction in the number of disordered atoms. Threading arms of the dislocation half-loop still contained a high number of disordered atoms. As the half-loop expanded sideways, more Ge atoms from the film moved towards the surface through the threading arms, further releasing strain. This caused an additional increase in the number of vacancies in the region, although at a slightly slower pace compared to stage I. 
\item
Stage III corresponds to the situation where the final 90$^{\circ}$ MD dislocation is already formed. The number of disordered atoms restores to the typical value of the film surface within the region, and the final number of vacancies correlates to the number of atoms of the missing plane above the dislocation.
\end{itemize}

Given the relevance of disordered regions in the origin and propagation of dislocations, we characterized their nature. In particular, we computed the coordination number and the radial distribution function, $g(r)$, of atoms in the disordered region after depositing 36 Ge MLs at 900 K (the same sample shown in Fig.~\ref{fig:nucleation} at t=415 ns, corresponding to the maximum size of the disordered region).

\begin{figure}[htb!]
    \centering
    \includegraphics[width=\columnwidth]{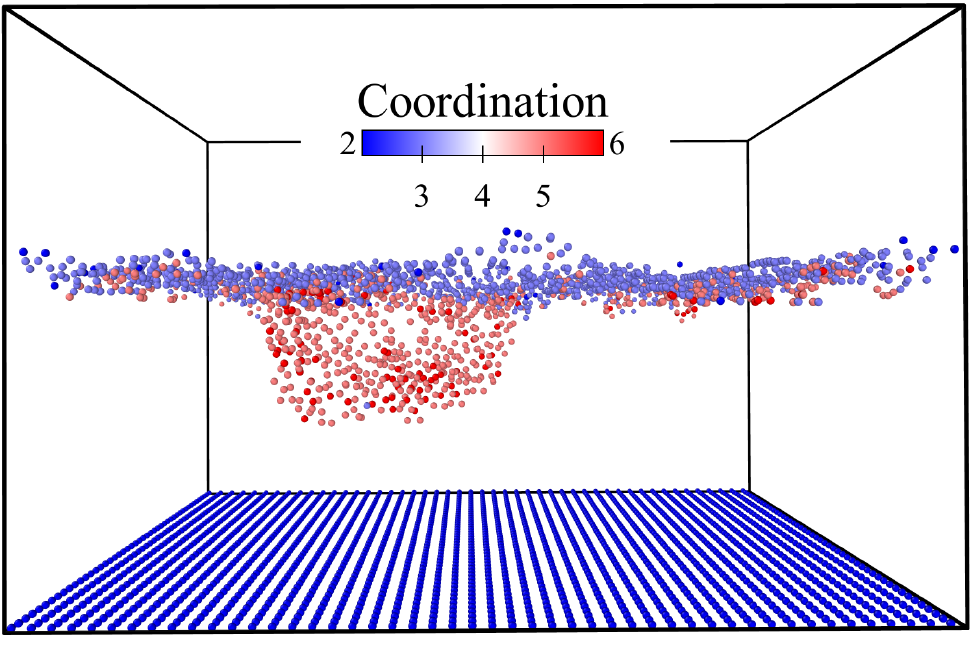}
    \caption{Perspective view of the sample after depositing 36 Ge MLs at 900 K, where atoms are colored according to their CN. Only atoms with CN different from 4 are shown.}
    \label{fig:cn}
\end{figure}

The coordination number (CN) of a given atom is defined as the number of nearest-neighbor atoms surrounding it. In our simulations, we defined nearest neighbors as atoms separated by a distance of less than 3.1$\textup{~\AA}$. This choice was based on the distance where $g(r)$ approaches zero beyond its first peak (i.e. the distance at which nearest neighbors are commonly found in the perfect crystal). Figure~\ref{fig:cn} shows atoms with with a coordination number different from four, in a perspective view of the analyzed sample. Atoms on the surface were under-coordinated, while atoms with CN=5 were abundant in the disordered region, and even some atoms with CN=6 were identified. The shape of this region with over-coordinated atoms resembles the one of Figure~\ref{fig:nucleation} at $t$=415 ns. However, they are not identical since fewer atoms are displayed here (many atoms classified as cubic diamond up to \nth{1} or \nth{2} neighbors and some of the \textit{other} atoms have CN=4). We found that approximately 80\% of the atoms in the disordered region had CN=4, 15\% had CN=5, and 5\% had CN=6, which is typical of the amorphous Ge phase, $a$-Ge~\cite{Bording2000,Lai2017}.

\begin{figure}[htb!]
    \centering
    \includegraphics[width=\columnwidth]{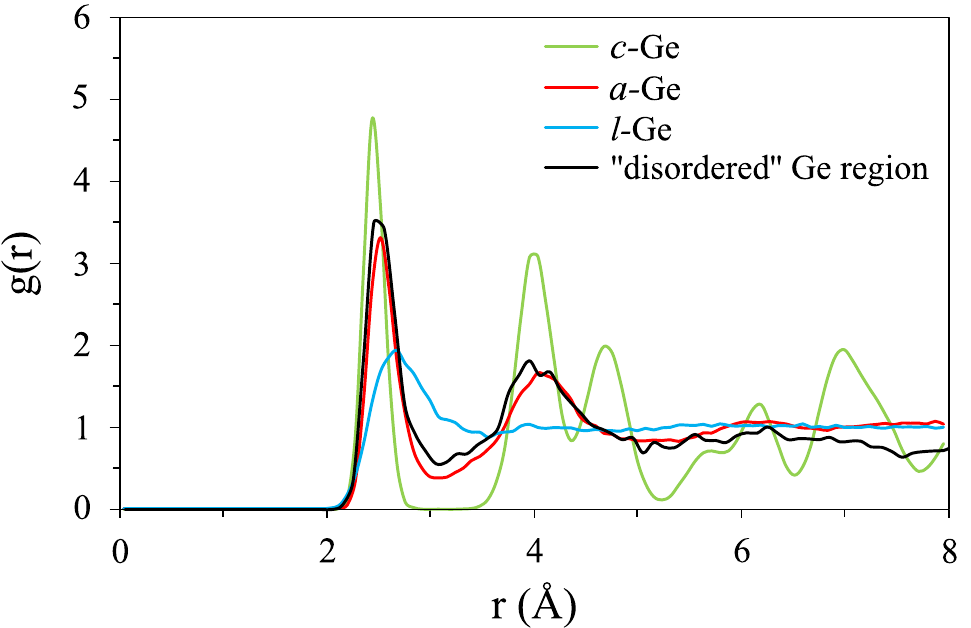}
    \caption{Radial distribution functions of the disordered region shown in Figure~\ref{fig:nucleation}, and of ideal $a$-Ge (at 900 K), $c$-Ge (at 900 K) and $l$-Ge (at 3000 K) samples.}
    \label{fig:gr}
\end{figure}

Figure~\ref{fig:gr} compares the $g(r)$ of the disordered region identified by non-cubic diamond atoms together with the $g(r)$ extracted from ideal $a$-Ge (at 900 K), $c$-Ge (at 900 K) and $l$-Ge (at 3000 K) samples. The $g(r)$ of the disordered region is very similar to that of ideal $a$-Ge (it does not asymptotically go to one with $r$ due to the limited volume used for its calculation), which confirms its amorphous phase nature. Therefore, the disordered structure identified prior to the dislocation formation is compatible with the $a$-Ge phase. 

We also quantified the atomic local density in a parallelepiped volume encompassing the disordered structure (excluding surface atoms). In a coherently grown Ge thin film on Si, this volume should contain a total of 1550 atoms, but we found that 162 atoms were missing, which corresponds to a local atomic density $\sim$ 10\% below that of the Ge crystal phase, $c$-Ge. We also analyzed the disordered regions that preceded dislocations in the rest of our simulations, and we found that missing atoms corresponded to local atomic densities ranging from 5\% to 15\% below that of $c$-Ge. Since ideal $a$-Ge has a local density only 5\% lower than $c$-Ge, our results indicate that disordered regions actually correspond to a low-density $a$-Ge phase.

\section{Conclusions}\label{sec::conclussions}

We studied the heteroepitaxial growth of Ge on Si(001) 2$\times$1 using CMD simulations with the SW potential, paying special attention to dislocation formation. We showed that deposition rates feasible in CMD (from  10$^6$ to 10$^8$ MLs/s) at elevated temperatures (between 700 and 1100 K) allow the simulation of the growth process in conditions analogous to those of experiments in out-of-equilibrium conditions (high deposition rates and low temperatures), where strain relaxation occurs through dislocation nucleation. In particular, we found that CMD simulations with deposition rates of 10$^8$ MLs/s and temperatures of 900 and 1000 K offer a good compromise between computational cost and accuracy. We tested that the features of the simulated grown film (critical thickness, intermixing, surface roughness) and the dislocation typology are in agreement with experiments.

We analyzed, at atomic level, the origin of dislocations and showed the interplay between surface roughness and dislocations. We observed that dislocations nucleated preferentially in valleys among islands on the Ge film surface, although the exact nucleation region where they appeared was, in principle, unpredictable. In all cases, regions where dislocations nucleated had low atomic density and were of amorphous nature. These amorphous regions grew as deposition proceeded until they reached a critical size. The high strain in the surface valleys, along with the low density and amorphous nature of these regions, favored a local increase of the atomic mobility. Some of their atoms moved towards the surface (generation of “vacancies”), thus releasing the stress accumulated in the growing Ge film. At some point, these regions recrystallized and propagated through the Ge film with the shape of a half-loop dislocation. The rough surface also played a role in the propagation of the dislocation half-loop as its ends moved along the valleys avoiding islands on the surface, forcing the dislocation line to change directions or even splitting. In turn, the formation of dislocations modified the surface roughness as the climbing atoms filled up the valleys and the strain released by the dislocation formation favored smoother surfaces during further deposition.

In our simulations with the SW potential at lower temperatures (650 and 800 K), even though dislocations were in an early stage of development, we also observed the formation of amorphous regions prior to the generation of the dislocation half-loops. We also tested deposition simulations with Tersoff potential and obtained analogous results concerning the amorphous nature of the nucleation region and its evolution. These findings indicate that the described mechanism is not related to the specific empirical potential, nor to the high temperatures used in the simulations, but it is an intrinsic feature of the early stages of dislocation formation in stressed and rough Ge films. 

\section*{Acknowledgments}
This work has been supported by the Spanish Ministerio de Ciencia e Innovación under Project No. PID2020-115118GB-I00.


\begin{appendix}

\section{Terminology used for the atomic identification of crystal structures}\label{appendixA}

Following OVITO's terminology, we classify atoms in our CMD simulations as:

\begin{itemize}
\item 
Cubic diamond: atom that has all of its \nth{1} and \nth{2} nearest neighbors positioned on cubic
diamond lattice sites.
\item
Cubic diamond up to \nth{1} neighbors: atom that is a \nth{1} neighbor of an atom that was classified as
“cubic diamond”. Its four nearest neighbors are positioned on lattice sites, but at least
one of its \nth{2} nearest neighbors is not.
\item
Cubic diamond up to \nth{2} neighbors: atom that is a \nth{2} nearest neighbor of an atom that was
classified as “cubic diamond”. The atom itself is positioned on a lattice site, but at least
one of its neighbors is missing or is not positioned on a lattice site.
\item
Hexagonal diamond: here we encompass atoms that have all of its \nth{1} and \nth{2} nearest neighbors positioned on hexagonal diamond lattice sites, and also those that are \nth{1} or \nth{2} neighbors to them.
\item
\textit{Other}: atom, with unknown coordination structure, that does not belong to any of the
previous categories. This atom has a disordered local environment, and it is typically located
inside or close to lattice defects (e.g. vacancies, interstitials, dislocations, ...), the surface
or an amorphous region.

\end{itemize}
\end{appendix}

\bibliography{biblio.bib}

\begin{thebibliography}{48}%
\makeatletter
\providecommand \@ifxundefined [1]{%
 \@ifx{#1\undefined}
}%
\providecommand \@ifnum [1]{%
 \ifnum #1\expandafter \@firstoftwo
 \else \expandafter \@secondoftwo
 \fi
}%
\providecommand \@ifx [1]{%
 \ifx #1\expandafter \@firstoftwo
 \else \expandafter \@secondoftwo
 \fi
}%
\providecommand \natexlab [1]{#1}%
\providecommand \enquote  [1]{``#1''}%
\providecommand \bibnamefont  [1]{#1}%
\providecommand \bibfnamefont [1]{#1}%
\providecommand \citenamefont [1]{#1}%
\providecommand \href@noop [0]{\@secondoftwo}%
\providecommand \href [0]{\begingroup \@sanitize@url \@href}%
\providecommand \@href[1]{\@@startlink{#1}\@@href}%
\providecommand \@@href[1]{\endgroup#1\@@endlink}%
\providecommand \@sanitize@url [0]{\catcode `\\12\catcode `\$12\catcode `\&12\catcode `\#12\catcode `\^12\catcode `\_12\catcode `\%12\relax}%
\providecommand \@@startlink[1]{}%
\providecommand \@@endlink[0]{}%
\providecommand \url  [0]{\begingroup\@sanitize@url \@url }%
\providecommand \@url [1]{\endgroup\@href {#1}{\urlprefix }}%
\providecommand \urlprefix  [0]{URL }%
\providecommand \Eprint [0]{\href }%
\providecommand \doibase [0]{https://doi.org/}%
\providecommand \selectlanguage [0]{\@gobble}%
\providecommand \bibinfo  [0]{\@secondoftwo}%
\providecommand \bibfield  [0]{\@secondoftwo}%
\providecommand \translation [1]{[#1]}%
\providecommand \BibitemOpen [0]{}%
\providecommand \bibitemStop [0]{}%
\providecommand \bibitemNoStop [0]{.\EOS\space}%
\providecommand \EOS [0]{\spacefactor3000\relax}%
\providecommand \BibitemShut  [1]{\csname bibitem#1\endcsname}%
\let\auto@bib@innerbib\@empty
\bibitem [{\citenamefont {Ryzhak}\ \emph {et~al.}(2024)\citenamefont {Ryzhak}, \citenamefont {Aberl}, \citenamefont {Prado-Navarrete}, \citenamefont {Vukušić}, \citenamefont {Corley-Wiciak}, \citenamefont {Skibitzki}, \citenamefont {Zoellner}, \citenamefont {Schubert}, \citenamefont {Virgilio}, \citenamefont {Brehm}, \citenamefont {Capellini},\ and\ \citenamefont {Spirito}}]{Ryzhak2024}%
  \BibitemOpen
  \bibfield  {author} {\bibinfo {author} {\bibfnamefont {D.}~\bibnamefont {Ryzhak}}, \bibinfo {author} {\bibfnamefont {J.}~\bibnamefont {Aberl}}, \bibinfo {author} {\bibfnamefont {E.}~\bibnamefont {Prado-Navarrete}}, \bibinfo {author} {\bibfnamefont {L.}~\bibnamefont {Vukušić}}, \bibinfo {author} {\bibfnamefont {A.~A.}\ \bibnamefont {Corley-Wiciak}}, \bibinfo {author} {\bibfnamefont {O.}~\bibnamefont {Skibitzki}}, \bibinfo {author} {\bibfnamefont {M.~H.}\ \bibnamefont {Zoellner}}, \bibinfo {author} {\bibfnamefont {M.~A.}\ \bibnamefont {Schubert}}, \bibinfo {author} {\bibfnamefont {M.}~\bibnamefont {Virgilio}}, \bibinfo {author} {\bibfnamefont {M.}~\bibnamefont {Brehm}}, \bibinfo {author} {\bibfnamefont {G.}~\bibnamefont {Capellini}},\ and\ \bibinfo {author} {\bibfnamefont {D.}~\bibnamefont {Spirito}},\ }\bibfield  {title} {\bibinfo {title} {Nanoheteroepitaxy of ge and sige on si: role of composition and capping on quantum dot photoluminescence},\ }\href {https://doi.org/10.1088/1361-6528/ad7f5f} {\bibfield
  {journal} {\bibinfo  {journal} {Nanotechnology}\ }\textbf {\bibinfo {volume} {35}},\ \bibinfo {pages} {505001} (\bibinfo {year} {2024})}\BibitemShut {NoStop}%
\bibitem [{\citenamefont {Arapkina}\ \emph {et~al.}(2023)\citenamefont {Arapkina}, \citenamefont {Chizh}, \citenamefont {Dubkov}, \citenamefont {Storozhevykh},\ and\ \citenamefont {Yuryev}}]{ARAPKINA2023}%
  \BibitemOpen
  \bibfield  {author} {\bibinfo {author} {\bibfnamefont {L.~V.}\ \bibnamefont {Arapkina}}, \bibinfo {author} {\bibfnamefont {K.~V.}\ \bibnamefont {Chizh}}, \bibinfo {author} {\bibfnamefont {V.~P.}\ \bibnamefont {Dubkov}}, \bibinfo {author} {\bibfnamefont {M.~S.}\ \bibnamefont {Storozhevykh}},\ and\ \bibinfo {author} {\bibfnamefont {V.~A.}\ \bibnamefont {Yuryev}},\ }\bibfield  {title} {\bibinfo {title} {Evolution of {Ge} wetting layers growing on smooth and rough {Si}(001) surfaces: {I}solated {105} facets as a kinetic factor of stress relaxation},\ }\href {https://doi.org/https://doi.org/10.1016/j.apsusc.2022.155094} {\bibfield  {journal} {\bibinfo  {journal} {Applied Surface Science}\ }\textbf {\bibinfo {volume} {608}},\ \bibinfo {pages} {155094} (\bibinfo {year} {2023})}\BibitemShut {NoStop}%
\bibitem [{\citenamefont {Du}\ \emph {et~al.}(2021)\citenamefont {Du}, \citenamefont {Kong}, \citenamefont {Toprak}, \citenamefont {Wang}, \citenamefont {Miao}, \citenamefont {Xu}, \citenamefont {Yu}, \citenamefont {Li}, \citenamefont {Lin}, \citenamefont {Han}, \citenamefont {Dong}, \citenamefont {Wang},\ and\ \citenamefont {Radamson}}]{Du2021}%
  \BibitemOpen
  \bibfield  {author} {\bibinfo {author} {\bibfnamefont {Y.}~\bibnamefont {Du}}, \bibinfo {author} {\bibfnamefont {Z.}~\bibnamefont {Kong}}, \bibinfo {author} {\bibfnamefont {M.~S.}\ \bibnamefont {Toprak}}, \bibinfo {author} {\bibfnamefont {G.}~\bibnamefont {Wang}}, \bibinfo {author} {\bibfnamefont {Y.}~\bibnamefont {Miao}}, \bibinfo {author} {\bibfnamefont {B.}~\bibnamefont {Xu}}, \bibinfo {author} {\bibfnamefont {J.}~\bibnamefont {Yu}}, \bibinfo {author} {\bibfnamefont {B.}~\bibnamefont {Li}}, \bibinfo {author} {\bibfnamefont {H.}~\bibnamefont {Lin}}, \bibinfo {author} {\bibfnamefont {J.}~\bibnamefont {Han}}, \bibinfo {author} {\bibfnamefont {Y.}~\bibnamefont {Dong}}, \bibinfo {author} {\bibfnamefont {W.}~\bibnamefont {Wang}},\ and\ \bibinfo {author} {\bibfnamefont {H.~H.}\ \bibnamefont {Radamson}},\ }\bibfield  {title} {\bibinfo {title} {Investigation of the heteroepitaxial process optimization of ge layers on si (001) by rpcvd},\ }\href {https://doi.org/10.3390/nano11040928} {\bibfield  {journal}
  {\bibinfo  {journal} {Nanomaterials}\ }\textbf {\bibinfo {volume} {11}},\ \bibinfo {pages} {928} (\bibinfo {year} {2021})}\BibitemShut {NoStop}%
\bibitem [{\citenamefont {Bolkhovityanov}\ \emph {et~al.}(2015)\citenamefont {Bolkhovityanov}, \citenamefont {Gutakovskii}, \citenamefont {Deryabin},\ and\ \citenamefont {Sokolov}}]{BOLKHOVITYANOV2015}%
  \BibitemOpen
  \bibfield  {author} {\bibinfo {author} {\bibfnamefont {Y.}~\bibnamefont {Bolkhovityanov}}, \bibinfo {author} {\bibfnamefont {A.}~\bibnamefont {Gutakovskii}}, \bibinfo {author} {\bibfnamefont {A.}~\bibnamefont {Deryabin}},\ and\ \bibinfo {author} {\bibfnamefont {L.}~\bibnamefont {Sokolov}},\ }\bibfield  {title} {\bibinfo {title} {Role of edge dislocations in plastic relaxation of gesi/si(001) heteroestructures: dependence of introduction mechanisms on film thickness},\ }\href {https://doi.org/https://doi.org/10.1134/S1063783415040071} {\bibfield  {journal} {\bibinfo  {journal} {Physics of the Solid State}\ }\textbf {\bibinfo {volume} {57}},\ \bibinfo {pages} {765} (\bibinfo {year} {2015})}\BibitemShut {NoStop}%
\bibitem [{\citenamefont {Bolkhovityanov}\ and\ \citenamefont {Sokolov}(2012)}]{BOLKHOVITYANOV2012_1}%
  \BibitemOpen
  \bibfield  {author} {\bibinfo {author} {\bibfnamefont {Y.~B.}\ \bibnamefont {Bolkhovityanov}}\ and\ \bibinfo {author} {\bibfnamefont {L.~V.}\ \bibnamefont {Sokolov}},\ }\bibfield  {title} {\bibinfo {title} {Ge-on-{S}i films obtained by epitaxial growing: edge dislocations and their participation in plastic relaxation},\ }\href {https://doi.org/10.1088/0268-1242/27/4/043001} {\bibfield  {journal} {\bibinfo  {journal} {Semiconductor Science and Technology}\ }\textbf {\bibinfo {volume} {27}},\ \bibinfo {pages} {043001} (\bibinfo {year} {2012})}\BibitemShut {NoStop}%
\bibitem [{\citenamefont {Wietler}\ \emph {et~al.}(2008)\citenamefont {Wietler}, \citenamefont {Bugiel},\ and\ \citenamefont {Hofmann}}]{WIETLER2008}%
  \BibitemOpen
  \bibfield  {author} {\bibinfo {author} {\bibfnamefont {T.~F.}\ \bibnamefont {Wietler}}, \bibinfo {author} {\bibfnamefont {E.}~\bibnamefont {Bugiel}},\ and\ \bibinfo {author} {\bibfnamefont {K.~R.}\ \bibnamefont {Hofmann}},\ }\bibfield  {title} {\bibinfo {title} {Relaxed germanium films on silicon (110)},\ }\href {https://doi.org/https://doi.org/10.1016/j.tsf.2008.08.018} {\bibfield  {journal} {\bibinfo  {journal} {Thin Solid Films}\ }\textbf {\bibinfo {volume} {517}},\ \bibinfo {pages} {272} (\bibinfo {year} {2008})}\BibitemShut {NoStop}%
\bibitem [{\citenamefont {Bolkhovityanov}\ \emph {et~al.}(2004)\citenamefont {Bolkhovityanov}, \citenamefont {Deryabin}, \citenamefont {Gutakovskii}, \citenamefont {Revenko},\ and\ \citenamefont {Sokolov}}]{bolkhovityanov2004}%
  \BibitemOpen
  \bibfield  {author} {\bibinfo {author} {\bibfnamefont {Y.~B.}\ \bibnamefont {Bolkhovityanov}}, \bibinfo {author} {\bibfnamefont {A.~S.}\ \bibnamefont {Deryabin}}, \bibinfo {author} {\bibfnamefont {A.~K.}\ \bibnamefont {Gutakovskii}}, \bibinfo {author} {\bibfnamefont {M.~A.}\ \bibnamefont {Revenko}},\ and\ \bibinfo {author} {\bibfnamefont {L.~V.}\ \bibnamefont {Sokolov}},\ }\bibfield  {title} {\bibinfo {title} {Direct observations of dislocation half-loops inserted from the surface of the {GeSi} heteroepitaxial film},\ }\href {https://doi.org/10.1063/1.1839271} {\bibfield  {journal} {\bibinfo  {journal} {Applied Physics Letters}\ }\textbf {\bibinfo {volume} {85}},\ \bibinfo {pages} {6140} (\bibinfo {year} {2004})}\BibitemShut {NoStop}%
\bibitem [{\citenamefont {Jesson}\ \emph {et~al.}(1993)\citenamefont {Jesson}, \citenamefont {Pennycook}, \citenamefont {Baribeau},\ and\ \citenamefont {Houghton}}]{Jesson1993}%
  \BibitemOpen
  \bibfield  {author} {\bibinfo {author} {\bibfnamefont {D.}~\bibnamefont {Jesson}}, \bibinfo {author} {\bibfnamefont {S.}~\bibnamefont {Pennycook}}, \bibinfo {author} {\bibfnamefont {J.-M.}\ \bibnamefont {Baribeau}},\ and\ \bibinfo {author} {\bibfnamefont {D.}~\bibnamefont {Houghton}},\ }\bibfield  {title} {\bibinfo {title} {Direct imaging of surface cusp evolution during strained-layer epitaxy and implications for strain relaxation},\ }\href {https://doi.org/https://doi.org/10.1103/PhysRevLett.71.1744} {\bibfield  {journal} {\bibinfo  {journal} {Physical Review Letters}\ }\textbf {\bibinfo {volume} {71}},\ \bibinfo {pages} {1744} (\bibinfo {year} {1993})}\BibitemShut {NoStop}%
\bibitem [{\citenamefont {Tersoff}\ and\ \citenamefont {LeGoues}(1994)}]{tersoff1994}%
  \BibitemOpen
  \bibfield  {author} {\bibinfo {author} {\bibfnamefont {J.}~\bibnamefont {Tersoff}}\ and\ \bibinfo {author} {\bibfnamefont {F.~K.}\ \bibnamefont {LeGoues}},\ }\bibfield  {title} {\bibinfo {title} {Competing relaxation mechanisms in strained layers},\ }\href {https://doi.org/10.1103/PhysRevLett.72.3570} {\bibfield  {journal} {\bibinfo  {journal} {Physical Review Letters}\ }\textbf {\bibinfo {volume} {72}},\ \bibinfo {pages} {3570} (\bibinfo {year} {1994})}\BibitemShut {NoStop}%
\bibitem [{\citenamefont {Pichaud}\ \emph {et~al.}(2009)\citenamefont {Pichaud}, \citenamefont {Burle}, \citenamefont {Texier}, \citenamefont {Alfonso}, \citenamefont {Gailhanou}, \citenamefont {Thibault-Pénisson}, \citenamefont {Fontaine},\ and\ \citenamefont {Vdovin}}]{Pichaud2009}%
  \BibitemOpen
  \bibfield  {author} {\bibinfo {author} {\bibfnamefont {B.}~\bibnamefont {Pichaud}}, \bibinfo {author} {\bibfnamefont {N.}~\bibnamefont {Burle}}, \bibinfo {author} {\bibfnamefont {M.}~\bibnamefont {Texier}}, \bibinfo {author} {\bibfnamefont {C.}~\bibnamefont {Alfonso}}, \bibinfo {author} {\bibfnamefont {M.}~\bibnamefont {Gailhanou}}, \bibinfo {author} {\bibfnamefont {J.}~\bibnamefont {Thibault-Pénisson}}, \bibinfo {author} {\bibfnamefont {C.}~\bibnamefont {Fontaine}},\ and\ \bibinfo {author} {\bibfnamefont {V.~I.}\ \bibnamefont {Vdovin}},\ }\bibfield  {title} {\bibinfo {title} {Dislocation nucleation in heteroepitaxial semiconducting films},\ }\href {https://doi.org/https://doi.org/10.1002/pssc.200881469} {\bibfield  {journal} {\bibinfo  {journal} {Physica status solidi c}\ }\textbf {\bibinfo {volume} {6}},\ \bibinfo {pages} {1827} (\bibinfo {year} {2009})}\BibitemShut {NoStop}%
\bibitem [{\citenamefont {Barbisan}\ \emph {et~al.}(2022)\citenamefont {Barbisan}, \citenamefont {Marzegalli},\ and\ \citenamefont {Montalenti}}]{Barbisan2022}%
  \BibitemOpen
  \bibfield  {author} {\bibinfo {author} {\bibfnamefont {L.}~\bibnamefont {Barbisan}}, \bibinfo {author} {\bibfnamefont {A.}~\bibnamefont {Marzegalli}},\ and\ \bibinfo {author} {\bibfnamefont {F.}~\bibnamefont {Montalenti}},\ }\bibfield  {title} {\bibinfo {title} {Atomic-scale insights on the formation of ordered arrays of edge dislocations in {G}e/{S}i(001) films via molecular dynamics simulations},\ }\href {https://doi.org/10.1038/s41598-022-07206-3} {\bibfield  {journal} {\bibinfo  {journal} {Scientific Reports}\ }\textbf {\bibinfo {volume} {12}},\ \bibinfo {pages} {3235} (\bibinfo {year} {2022})}\BibitemShut {NoStop}%
\bibitem [{\citenamefont {Capano}(1992)}]{Capano92}%
  \BibitemOpen
  \bibfield  {author} {\bibinfo {author} {\bibfnamefont {M.}~\bibnamefont {Capano}},\ }\bibfield  {title} {\bibinfo {title} {Multiplication of dislocations in si$_{1-x}$ge$_x$ layers on si(001)},\ }\href {https://doi.org/https://doi.org/10.1103/PhysRevB.45.11768} {\bibfield  {journal} {\bibinfo  {journal} {Physical Review B}\ }\textbf {\bibinfo {volume} {45}},\ \bibinfo {pages} {11768} (\bibinfo {year} {1992})}\BibitemShut {NoStop}%
\bibitem [{\citenamefont {Dong}\ \emph {et~al.}(1998)\citenamefont {Dong}, \citenamefont {Schnitker}, \citenamefont {Smith},\ and\ \citenamefont {Srolovitz}}]{dong1998}%
  \BibitemOpen
  \bibfield  {author} {\bibinfo {author} {\bibfnamefont {L.}~\bibnamefont {Dong}}, \bibinfo {author} {\bibfnamefont {J.}~\bibnamefont {Schnitker}}, \bibinfo {author} {\bibfnamefont {R.~W.}\ \bibnamefont {Smith}},\ and\ \bibinfo {author} {\bibfnamefont {D.~J.}\ \bibnamefont {Srolovitz}},\ }\bibfield  {title} {\bibinfo {title} {Stress relaxation and misfit dislocation nucleation in the growth of misfitting films: A molecular dynamics simulation study},\ }\href {https://doi.org/10.1063/1.366676} {\bibfield  {journal} {\bibinfo  {journal} {Journal of Applied Physics}\ }\textbf {\bibinfo {volume} {83}},\ \bibinfo {pages} {217} (\bibinfo {year} {1998})}\BibitemShut {NoStop}%
\bibitem [{\citenamefont {Maras}\ \emph {et~al.}(2016)\citenamefont {Maras}, \citenamefont {Trushin}, \citenamefont {Stukowski}, \citenamefont {Ala-Nissila},\ and\ \citenamefont {Jónsson}}]{maras}%
  \BibitemOpen
  \bibfield  {author} {\bibinfo {author} {\bibfnamefont {E.}~\bibnamefont {Maras}}, \bibinfo {author} {\bibfnamefont {O.}~\bibnamefont {Trushin}}, \bibinfo {author} {\bibfnamefont {A.}~\bibnamefont {Stukowski}}, \bibinfo {author} {\bibfnamefont {T.}~\bibnamefont {Ala-Nissila}},\ and\ \bibinfo {author} {\bibfnamefont {H.}~\bibnamefont {Jónsson}},\ }\bibfield  {title} {\bibinfo {title} {Global transition path search for dislocation formation in {G}e on {S}i(001)},\ }\href {https://doi.org/https://doi.org/10.1016/j.cpc.2016.04.001} {\bibfield  {journal} {\bibinfo  {journal} {Computer Physics Communications}\ }\textbf {\bibinfo {volume} {205}},\ \bibinfo {pages} {13} (\bibinfo {year} {2016})}\BibitemShut {NoStop}%
\bibitem [{\citenamefont {Maras}\ \emph {et~al.}(2017)\citenamefont {Maras}, \citenamefont {Pizzagalli}, \citenamefont {Ala-Nissila},\ and\ \citenamefont {J{\'o}nsson}}]{Maras2017}%
  \BibitemOpen
  \bibfield  {author} {\bibinfo {author} {\bibfnamefont {E.}~\bibnamefont {Maras}}, \bibinfo {author} {\bibfnamefont {L.}~\bibnamefont {Pizzagalli}}, \bibinfo {author} {\bibfnamefont {T.}~\bibnamefont {Ala-Nissila}},\ and\ \bibinfo {author} {\bibfnamefont {H.}~\bibnamefont {J{\'o}nsson}},\ }\bibfield  {title} {\bibinfo {title} {Atomic scale formation mechanism of edge dislocation relieving lattice strain in a {GeSi} overlayer on {S}i(001)},\ }\href {https://doi.org/10.1038/s41598-017-12009-y} {\bibfield  {journal} {\bibinfo  {journal} {Scientific Reports}\ }\textbf {\bibinfo {volume} {7}},\ \bibinfo {pages} {11966} (\bibinfo {year} {2017})}\BibitemShut {NoStop}%
\bibitem [{\citenamefont {Trushin}\ \emph {et~al.}(2016)\citenamefont {Trushin}, \citenamefont {Maras}, \citenamefont {Stukowski}, \citenamefont {Granato}, \citenamefont {Ying}, \citenamefont {Jónsson},\ and\ \citenamefont {Ala-Nissila}}]{Trushin_2016}%
  \BibitemOpen
  \bibfield  {author} {\bibinfo {author} {\bibfnamefont {O.}~\bibnamefont {Trushin}}, \bibinfo {author} {\bibfnamefont {E.}~\bibnamefont {Maras}}, \bibinfo {author} {\bibfnamefont {A.}~\bibnamefont {Stukowski}}, \bibinfo {author} {\bibfnamefont {E.}~\bibnamefont {Granato}}, \bibinfo {author} {\bibfnamefont {S.~C.}\ \bibnamefont {Ying}}, \bibinfo {author} {\bibfnamefont {H.}~\bibnamefont {Jónsson}},\ and\ \bibinfo {author} {\bibfnamefont {T.}~\bibnamefont {Ala-Nissila}},\ }\bibfield  {title} {\bibinfo {title} {Minimum energy path for the nucleation of misfit dislocations in {Ge/Si}(001) heteroepitaxy},\ }\href {https://doi.org/10.1088/0965-0393/24/3/035007} {\bibfield  {journal} {\bibinfo  {journal} {Modelling and Simulation in Materials Science and Engineering}\ }\textbf {\bibinfo {volume} {24}},\ \bibinfo {pages} {035007} (\bibinfo {year} {2016})}\BibitemShut {NoStop}%
\bibitem [{\citenamefont {Zhang}\ and\ \citenamefont {Cai}(2018)}]{zhang2018}%
  \BibitemOpen
  \bibfield  {author} {\bibinfo {author} {\bibfnamefont {X.}~\bibnamefont {Zhang}}\ and\ \bibinfo {author} {\bibfnamefont {W.}~\bibnamefont {Cai}},\ }\href@noop {} {\bibinfo {title} {Critical strain for surface nucleation of dislocations in silicon}} (\bibinfo {year} {2018}),\ \Eprint {https://arxiv.org/abs/1806.01974} {arXiv:1806.01974 [physics.app-ph]} \BibitemShut {NoStop}%
\bibitem [{\citenamefont {Shima}\ \emph {et~al.}(2010)\citenamefont {Shima}, \citenamefont {Izumi},\ and\ \citenamefont {Sakai}}]{izumi2010}%
  \BibitemOpen
  \bibfield  {author} {\bibinfo {author} {\bibfnamefont {K.}~\bibnamefont {Shima}}, \bibinfo {author} {\bibfnamefont {S.}~\bibnamefont {Izumi}},\ and\ \bibinfo {author} {\bibfnamefont {S.}~\bibnamefont {Sakai}},\ }\bibfield  {title} {\bibinfo {title} {Reaction pathway analysis for dislocation nucleation from a sharp corner in silicon: Glide set versus shuffle set},\ }\href {https://doi.org/10.1063/1.3486465} {\bibfield  {journal} {\bibinfo  {journal} {Journal of Applied Physics}\ }\textbf {\bibinfo {volume} {108}},\ \bibinfo {pages} {063504} (\bibinfo {year} {2010})}\BibitemShut {NoStop}%
\bibitem [{\citenamefont {Li}\ \emph {et~al.}(2012)\citenamefont {Li}, \citenamefont {Picu}, \citenamefont {Muralidhar},\ and\ \citenamefont {Oldiges}}]{Li2012}%
  \BibitemOpen
  \bibfield  {author} {\bibinfo {author} {\bibfnamefont {Z.}~\bibnamefont {Li}}, \bibinfo {author} {\bibfnamefont {R.~C.}\ \bibnamefont {Picu}}, \bibinfo {author} {\bibfnamefont {R.}~\bibnamefont {Muralidhar}},\ and\ \bibinfo {author} {\bibfnamefont {P.}~\bibnamefont {Oldiges}},\ }\bibfield  {title} {\bibinfo {title} {Effect of {G}e on dislocation nucleation from surface imperfections in {Si-Ge}},\ }\href {https://doi.org/10.1063/1.4745864} {\bibfield  {journal} {\bibinfo  {journal} {Journal of Applied Physics}\ }\textbf {\bibinfo {volume} {112}},\ \bibinfo {pages} {034315} (\bibinfo {year} {2012})}\BibitemShut {NoStop}%
\bibitem [{\citenamefont {Thompson}\ \emph {et~al.}(2022)\citenamefont {Thompson}, \citenamefont {Aktulga}, \citenamefont {Berger}, \citenamefont {Bolintineanu}, \citenamefont {Brown}, \citenamefont {Crozier}, \citenamefont {in~'t Veld}, \citenamefont {Kohlmeyer}, \citenamefont {Moore}, \citenamefont {Nguyen}, \citenamefont {Shan}, \citenamefont {Stevens}, \citenamefont {Tranchida}, \citenamefont {Trott},\ and\ \citenamefont {Plimpton}}]{lammps}%
  \BibitemOpen
  \bibfield  {author} {\bibinfo {author} {\bibfnamefont {A.~P.}\ \bibnamefont {Thompson}}, \bibinfo {author} {\bibfnamefont {H.~M.}\ \bibnamefont {Aktulga}}, \bibinfo {author} {\bibfnamefont {R.}~\bibnamefont {Berger}}, \bibinfo {author} {\bibfnamefont {D.~S.}\ \bibnamefont {Bolintineanu}}, \bibinfo {author} {\bibfnamefont {W.~M.}\ \bibnamefont {Brown}}, \bibinfo {author} {\bibfnamefont {P.~S.}\ \bibnamefont {Crozier}}, \bibinfo {author} {\bibfnamefont {P.~J.}\ \bibnamefont {in~'t Veld}}, \bibinfo {author} {\bibfnamefont {A.}~\bibnamefont {Kohlmeyer}}, \bibinfo {author} {\bibfnamefont {S.~G.}\ \bibnamefont {Moore}}, \bibinfo {author} {\bibfnamefont {T.~D.}\ \bibnamefont {Nguyen}}, \bibinfo {author} {\bibfnamefont {R.}~\bibnamefont {Shan}}, \bibinfo {author} {\bibfnamefont {M.~J.}\ \bibnamefont {Stevens}}, \bibinfo {author} {\bibfnamefont {J.}~\bibnamefont {Tranchida}}, \bibinfo {author} {\bibfnamefont {C.}~\bibnamefont {Trott}},\ and\ \bibinfo {author} {\bibfnamefont {S.~J.}\ \bibnamefont {Plimpton}},\
  }\bibfield  {title} {\bibinfo {title} {{LAMMPS} - a flexible simulation tool for particle-based materials modeling at the atomic, meso, and continuum scales},\ }\href {https://doi.org/10.1016/j.cpc.2021.108171} {\bibfield  {journal} {\bibinfo  {journal} {Comp. Phys. Comm.}\ }\textbf {\bibinfo {volume} {271}},\ \bibinfo {pages} {108171} (\bibinfo {year} {2022})}\BibitemShut {NoStop}%
\bibitem [{\citenamefont {Stukowski}(2009)}]{Stukowski_2010}%
  \BibitemOpen
  \bibfield  {author} {\bibinfo {author} {\bibfnamefont {A.}~\bibnamefont {Stukowski}},\ }\bibfield  {title} {\bibinfo {title} {Visualization and analysis of atomistic simulation data with {OVITO}–the {O}pen {V}isualization {T}ool},\ }\href {https://doi.org/10.1088/0965-0393/18/1/015012} {\bibfield  {journal} {\bibinfo  {journal} {Modelling and Simulation in Materials Science and Engineering}\ }\textbf {\bibinfo {volume} {18}},\ \bibinfo {pages} {015012} (\bibinfo {year} {2009})}\BibitemShut {NoStop}%
\bibitem [{\citenamefont {Balamane}\ \emph {et~al.}(1992)\citenamefont {Balamane}, \citenamefont {Halicioglu},\ and\ \citenamefont {Tiller}}]{balamane}%
  \BibitemOpen
  \bibfield  {author} {\bibinfo {author} {\bibfnamefont {H.}~\bibnamefont {Balamane}}, \bibinfo {author} {\bibfnamefont {T.}~\bibnamefont {Halicioglu}},\ and\ \bibinfo {author} {\bibfnamefont {W.~A.}\ \bibnamefont {Tiller}},\ }\bibfield  {title} {\bibinfo {title} {Comparative study of silicon empirical interatomic potentials},\ }\href {https://doi.org/https://doi.org/10.1103/PhysRevB.46.2250} {\bibfield  {journal} {\bibinfo  {journal} {Phys. Rev. B}\ }\textbf {\bibinfo {volume} {46}},\ \bibinfo {pages} {2250} (\bibinfo {year} {1992})}\BibitemShut {NoStop}%
\bibitem [{\citenamefont {Posselt}\ and\ \citenamefont {Gabriel}(2009)}]{swp}%
  \BibitemOpen
  \bibfield  {author} {\bibinfo {author} {\bibfnamefont {M.}~\bibnamefont {Posselt}}\ and\ \bibinfo {author} {\bibfnamefont {A.}~\bibnamefont {Gabriel}},\ }\bibfield  {title} {\bibinfo {title} {Atomistic simulation of amorphous germanium and its solid phase epitaxial recrystallization},\ }\href {https://doi.org/https://doi.org/10.1103/PhysRevB.80.045202} {\bibfield  {journal} {\bibinfo  {journal} {Phys. Rev. B}\ }\textbf {\bibinfo {volume} {80}},\ \bibinfo {pages} {045202} (\bibinfo {year} {2009})}\BibitemShut {NoStop}%
\bibitem [{\citenamefont {Grabow}\ and\ \citenamefont {Gilmer}(1988)}]{gilmergrabow}%
  \BibitemOpen
  \bibfield  {author} {\bibinfo {author} {\bibfnamefont {M.~H.}\ \bibnamefont {Grabow}}\ and\ \bibinfo {author} {\bibfnamefont {G.~H.}\ \bibnamefont {Gilmer}},\ }\bibfield  {title} {\bibinfo {title} {Thin film growth modes, wetting and cluster nucleation},\ }\href {https://doi.org/https://doi.org/10.1016/0039-6028(88)90858-8} {\bibfield  {journal} {\bibinfo  {journal} {Surface Science}\ }\textbf {\bibinfo {volume} {194}},\ \bibinfo {pages} {333} (\bibinfo {year} {1988})}\BibitemShut {NoStop}%
\bibitem [{\citenamefont {Martín-Encinar}\ \emph {et~al.}(2023)\citenamefont {Martín-Encinar}, \citenamefont {Marqués}, \citenamefont {Santos}, \citenamefont {López},\ and\ \citenamefont {Pelaz}}]{luis2023}%
  \BibitemOpen
  \bibfield  {author} {\bibinfo {author} {\bibfnamefont {L.}~\bibnamefont {Martín-Encinar}}, \bibinfo {author} {\bibfnamefont {L.~A.}\ \bibnamefont {Marqués}}, \bibinfo {author} {\bibfnamefont {I.}~\bibnamefont {Santos}}, \bibinfo {author} {\bibfnamefont {P.}~\bibnamefont {López}},\ and\ \bibinfo {author} {\bibfnamefont {L.}~\bibnamefont {Pelaz}},\ }\bibfield  {title} {\bibinfo {title} {Concurrent characterization of surface diffusion and intermixing of {G}e on {S}i: A {C}lassical {M}olecular {D}ynamics study},\ }\href {https://doi.org/https://doi.org/10.1002/adts.202200848} {\bibfield  {journal} {\bibinfo  {journal} {Advanced Theory and Simulations}\ }\textbf {\bibinfo {volume} {6}},\ \bibinfo {pages} {2200848} (\bibinfo {year} {2023})}\BibitemShut {NoStop}%
\bibitem [{\citenamefont {Carlsson}\ and\ \citenamefont {Martin}(2010)}]{CARLSSON2010314}%
  \BibitemOpen
  \bibfield  {author} {\bibinfo {author} {\bibfnamefont {J.-O.}\ \bibnamefont {Carlsson}}\ and\ \bibinfo {author} {\bibfnamefont {P.~M.}\ \bibnamefont {Martin}},\ }\bibfield  {title} {\bibinfo {title} {Handbook of deposition technologies for films and coatings (third edition)}\ }(\bibinfo  {publisher} {William Andrew Publishing},\ \bibinfo {year} {2010})\BibitemShut {NoStop}%
\bibitem [{\citenamefont {Frigeri}\ \emph {et~al.}(2011)\citenamefont {Frigeri}, \citenamefont {Seravalli}, \citenamefont {Trevisi},\ and\ \citenamefont {Franchi}}]{FRIGERI2011480}%
  \BibitemOpen
  \bibfield  {author} {\bibinfo {author} {\bibfnamefont {P.}~\bibnamefont {Frigeri}}, \bibinfo {author} {\bibfnamefont {L.}~\bibnamefont {Seravalli}}, \bibinfo {author} {\bibfnamefont {G.}~\bibnamefont {Trevisi}},\ and\ \bibinfo {author} {\bibfnamefont {S.}~\bibnamefont {Franchi}},\ }\bibfield  {title} {\bibinfo {title} {Comprehensive semiconductor science and technology}\ }(\bibinfo  {publisher} {Elsevier},\ \bibinfo {year} {2011})\BibitemShut {NoStop}%
\bibitem [{\citenamefont {Lagally}(1993)}]{lagally}%
  \BibitemOpen
  \bibfield  {author} {\bibinfo {author} {\bibfnamefont {M.~G.}\ \bibnamefont {Lagally}},\ }\bibfield  {title} {\bibinfo {title} {An atomic-level view of kinetic and thermodynamic influences in the growth of thin films},\ }\href {https://doi.org/https://doi.org/10.1143/jjap.32.1493} {\bibfield  {journal} {\bibinfo  {journal} {Japanese Journal of Applied Physics}\ }\textbf {\bibinfo {volume} {32}},\ \bibinfo {pages} {1493} (\bibinfo {year} {1993})}\BibitemShut {NoStop}%
\bibitem [{\citenamefont {Radić}\ \emph {et~al.}(2006)\citenamefont {Radić}, \citenamefont {Pivac}, \citenamefont {Dubček}, \citenamefont {Kovačević},\ and\ \citenamefont {Bernstorff}}]{RADIC2006}%
  \BibitemOpen
  \bibfield  {author} {\bibinfo {author} {\bibfnamefont {N.}~\bibnamefont {Radić}}, \bibinfo {author} {\bibfnamefont {B.}~\bibnamefont {Pivac}}, \bibinfo {author} {\bibfnamefont {P.}~\bibnamefont {Dubček}}, \bibinfo {author} {\bibfnamefont {I.}~\bibnamefont {Kovačević}},\ and\ \bibinfo {author} {\bibfnamefont {S.}~\bibnamefont {Bernstorff}},\ }\bibfield  {title} {\bibinfo {title} {Growth of {Ge} islands on {Si} substrates},\ }\href {https://doi.org/https://doi.org/10.1016/j.tsf.2005.12.198} {\bibfield  {journal} {\bibinfo  {journal} {Thin Solid Films}\ }\textbf {\bibinfo {volume} {515}},\ \bibinfo {pages} {752} (\bibinfo {year} {2006})}\BibitemShut {NoStop}%
\bibitem [{\citenamefont {Capellini}\ \emph {et~al.}(2003)\citenamefont {Capellini}, \citenamefont {De~Seta},\ and\ \citenamefont {Evangelisti}}]{Capellini2003}%
  \BibitemOpen
  \bibfield  {author} {\bibinfo {author} {\bibfnamefont {G.}~\bibnamefont {Capellini}}, \bibinfo {author} {\bibfnamefont {M.}~\bibnamefont {De~Seta}},\ and\ \bibinfo {author} {\bibfnamefont {F.}~\bibnamefont {Evangelisti}},\ }\bibfield  {title} {\bibinfo {title} {Ge/{S}i(100) islands: Growth dynamics versus growth rate},\ }\href {https://doi.org/10.1063/1.1527211} {\bibfield  {journal} {\bibinfo  {journal} {Journal of Applied Physics}\ }\textbf {\bibinfo {volume} {93}},\ \bibinfo {pages} {291} (\bibinfo {year} {2003})}\BibitemShut {NoStop}%
\bibitem [{\citenamefont {Choi}\ \emph {et~al.}(2008)\citenamefont {Choi}, \citenamefont {Ge}, \citenamefont {Harris}, \citenamefont {Cagnon},\ and\ \citenamefont {Stemmer}}]{CHOI2008}%
  \BibitemOpen
  \bibfield  {author} {\bibinfo {author} {\bibfnamefont {D.}~\bibnamefont {Choi}}, \bibinfo {author} {\bibfnamefont {Y.}~\bibnamefont {Ge}}, \bibinfo {author} {\bibfnamefont {J.~S.}\ \bibnamefont {Harris}}, \bibinfo {author} {\bibfnamefont {J.}~\bibnamefont {Cagnon}},\ and\ \bibinfo {author} {\bibfnamefont {S.}~\bibnamefont {Stemmer}},\ }\bibfield  {title} {\bibinfo {title} {Low surface roughness and threading dislocation density {Ge} growth on {Si} (001)},\ }\href {https://doi.org/https://doi.org/10.1016/j.jcrysgro.2008.07.029} {\bibfield  {journal} {\bibinfo  {journal} {Journal of Crystal Growth}\ }\textbf {\bibinfo {volume} {310}},\ \bibinfo {pages} {4273} (\bibinfo {year} {2008})}\BibitemShut {NoStop}%
\bibitem [{\citenamefont {Shah}\ \emph {et~al.}(2011)\citenamefont {Shah}, \citenamefont {Dobbie}, \citenamefont {Myronov},\ and\ \citenamefont {Leadley}}]{SHAH2011}%
  \BibitemOpen
  \bibfield  {author} {\bibinfo {author} {\bibfnamefont {V.}~\bibnamefont {Shah}}, \bibinfo {author} {\bibfnamefont {A.}~\bibnamefont {Dobbie}}, \bibinfo {author} {\bibfnamefont {M.}~\bibnamefont {Myronov}},\ and\ \bibinfo {author} {\bibfnamefont {D.}~\bibnamefont {Leadley}},\ }\bibfield  {title} {\bibinfo {title} {High quality relaxed {Ge} layers grown directly on a {Si}(001) substrate},\ }\href {https://doi.org/https://doi.org/10.1016/j.sse.2011.03.005} {\bibfield  {journal} {\bibinfo  {journal} {Solid-State Electronics}\ }\textbf {\bibinfo {volume} {62}},\ \bibinfo {pages} {189} (\bibinfo {year} {2011})}\BibitemShut {NoStop}%
\bibitem [{\citenamefont {Ikeda}\ \emph {et~al.}(1997)\citenamefont {Ikeda}, \citenamefont {Sumitomo}, \citenamefont {Nishioka}, \citenamefont {Yasue}, \citenamefont {Koshikawa},\ and\ \citenamefont {Kido}}]{ikeda}%
  \BibitemOpen
  \bibfield  {author} {\bibinfo {author} {\bibfnamefont {A.}~\bibnamefont {Ikeda}}, \bibinfo {author} {\bibfnamefont {K.}~\bibnamefont {Sumitomo}}, \bibinfo {author} {\bibfnamefont {T.}~\bibnamefont {Nishioka}}, \bibinfo {author} {\bibfnamefont {T.}~\bibnamefont {Yasue}}, \bibinfo {author} {\bibfnamefont {T.}~\bibnamefont {Koshikawa}},\ and\ \bibinfo {author} {\bibfnamefont {Y.}~\bibnamefont {Kido}},\ }\bibfield  {title} {\bibinfo {title} {Intermixing at {Ge/Si}(001) interfaces studied by surface energy loss of medium energy ion scattering},\ }\href {https://doi.org/https://doi.org/10.1016/S0039-6028(97)00275-6} {\bibfield  {journal} {\bibinfo  {journal} {Surface Science}\ }\textbf {\bibinfo {volume} {385}},\ \bibinfo {pages} {200} (\bibinfo {year} {1997})}\BibitemShut {NoStop}%
\bibitem [{\citenamefont {Cho}\ and\ \citenamefont {Kang}(2000)}]{cho}%
  \BibitemOpen
  \bibfield  {author} {\bibinfo {author} {\bibfnamefont {J.-H.}\ \bibnamefont {Cho}}\ and\ \bibinfo {author} {\bibfnamefont {M.-H.}\ \bibnamefont {Kang}},\ }\bibfield  {title} {\bibinfo {title} {Ge-{S}i intermixing at the {Ge/Si}(001) surface},\ }\href {https://doi.org/https://doi.org/10.1103/PhysRevB.61.1688} {\bibfield  {journal} {\bibinfo  {journal} {Phys. Rev. B}\ }\textbf {\bibinfo {volume} {61}},\ \bibinfo {pages} {1688} (\bibinfo {year} {2000})}\BibitemShut {NoStop}%
\bibitem [{\citenamefont {Wagner}\ and\ \citenamefont {Gulari}(2004)}]{wagner}%
  \BibitemOpen
  \bibfield  {author} {\bibinfo {author} {\bibfnamefont {R.~J.}\ \bibnamefont {Wagner}}\ and\ \bibinfo {author} {\bibfnamefont {E.}~\bibnamefont {Gulari}},\ }\bibfield  {title} {\bibinfo {title} {Simulation of {Ge/Si} intermixing during heteroepitaxy},\ }\href {https://doi.org/https://doi.org/10.1103/PhysRevB.69.195312} {\bibfield  {journal} {\bibinfo  {journal} {Phys. Rev. B}\ }\textbf {\bibinfo {volume} {69}},\ \bibinfo {pages} {195312} (\bibinfo {year} {2004})}\BibitemShut {NoStop}%
\bibitem [{\citenamefont {Arroyo}\ \emph {et~al.}(2019)\citenamefont {Arroyo}, \citenamefont {Isa}, \citenamefont {Isella}, \citenamefont {Erni}, \citenamefont {{von Känel}}, \citenamefont {Gröning},\ and\ \citenamefont {Rossell}}]{ARROYO2019}%
  \BibitemOpen
  \bibfield  {author} {\bibinfo {author} {\bibfnamefont {R.~D.}\ \bibnamefont {Arroyo}}, \bibinfo {author} {\bibfnamefont {F.}~\bibnamefont {Isa}}, \bibinfo {author} {\bibfnamefont {G.}~\bibnamefont {Isella}}, \bibinfo {author} {\bibfnamefont {R.}~\bibnamefont {Erni}}, \bibinfo {author} {\bibfnamefont {H.}~\bibnamefont {{von Känel}}}, \bibinfo {author} {\bibfnamefont {P.}~\bibnamefont {Gröning}},\ and\ \bibinfo {author} {\bibfnamefont {M.~D.}\ \bibnamefont {Rossell}},\ }\bibfield  {title} {\bibinfo {title} {Effect of thermal annealing on the interface quality of {Ge/Si} heterostructures},\ }\href {https://doi.org/https://doi.org/10.1016/j.scriptamat.2019.05.025} {\bibfield  {journal} {\bibinfo  {journal} {Scripta Materialia}\ }\textbf {\bibinfo {volume} {170}},\ \bibinfo {pages} {52} (\bibinfo {year} {2019})}\BibitemShut {NoStop}%
\bibitem [{\citenamefont {Matthews}(1975)}]{mattews1975}%
  \BibitemOpen
  \bibfield  {author} {\bibinfo {author} {\bibfnamefont {J.~W.}\ \bibnamefont {Matthews}},\ }\bibfield  {title} {\bibinfo {title} {{Defects associated with the accommodation of misfit between crystals}},\ }\href {https://doi.org/10.1116/1.568741} {\bibfield  {journal} {\bibinfo  {journal} {Journal of Vacuum Science and Technology}\ }\textbf {\bibinfo {volume} {12}},\ \bibinfo {pages} {126} (\bibinfo {year} {1975})}\BibitemShut {NoStop}%
\bibitem [{\citenamefont {People}\ and\ \citenamefont {Bean}(1985)}]{People1985}%
  \BibitemOpen
  \bibfield  {author} {\bibinfo {author} {\bibfnamefont {R.}~\bibnamefont {People}}\ and\ \bibinfo {author} {\bibfnamefont {J.~C.}\ \bibnamefont {Bean}},\ }\bibfield  {title} {\bibinfo {title} {Calculation of critical layer thickness versus lattice mismatch for {G}e$_x${S}i$_{1-x}$/{S}i strained‐layer heterostructures},\ }\href {https://doi.org/10.1063/1.96206} {\bibfield  {journal} {\bibinfo  {journal} {Applied Physics Letters}\ }\textbf {\bibinfo {volume} {47}},\ \bibinfo {pages} {322} (\bibinfo {year} {1985})}\BibitemShut {NoStop}%
\bibitem [{\citenamefont {Bean}\ \emph {et~al.}(1984)\citenamefont {Bean}, \citenamefont {Feldman}, \citenamefont {Fiory}, \citenamefont {Nakahara},\ and\ \citenamefont {Robinson}}]{bean1984}%
  \BibitemOpen
  \bibfield  {author} {\bibinfo {author} {\bibfnamefont {J.}~\bibnamefont {Bean}}, \bibinfo {author} {\bibfnamefont {L.~C.}\ \bibnamefont {Feldman}}, \bibinfo {author} {\bibfnamefont {A.}~\bibnamefont {Fiory}}, \bibinfo {author} {\bibfnamefont {S.~t.}\ \bibnamefont {Nakahara}},\ and\ \bibinfo {author} {\bibfnamefont {I.}~\bibnamefont {Robinson}},\ }\bibfield  {title} {\bibinfo {title} {Ge$_x${S}i$_{1-x}$/{S}i strained-layer superlattice grown by molecular beam epitaxy},\ }\href {https://doi.org/10.1116/1.572361} {\bibfield  {journal} {\bibinfo  {journal} {Journal of Vacuum Science and Technology A: Vacuum, Surfaces, and Films}\ }\textbf {\bibinfo {volume} {2}},\ \bibinfo {pages} {436} (\bibinfo {year} {1984})}\BibitemShut {NoStop}%
\bibitem [{\citenamefont {Houghton}\ \emph {et~al.}(1990)\citenamefont {Houghton}, \citenamefont {Perovic}, \citenamefont {Baribeau},\ and\ \citenamefont {Weatherly}}]{houghton1990m}%
  \BibitemOpen
  \bibfield  {author} {\bibinfo {author} {\bibfnamefont {D.~C.}\ \bibnamefont {Houghton}}, \bibinfo {author} {\bibfnamefont {D.~D.}\ \bibnamefont {Perovic}}, \bibinfo {author} {\bibfnamefont {J.}~\bibnamefont {Baribeau}},\ and\ \bibinfo {author} {\bibfnamefont {G.~C.}\ \bibnamefont {Weatherly}},\ }\bibfield  {title} {\bibinfo {title} {Misfit strain relaxation in {G}e$_{x}${S}i$_{1-x}$/{S}i heterostructures: The structural stability of buried strained layers and strained‐layer superlattices},\ }\href {https://doi.org/10.1063/1.345613} {\bibfield  {journal} {\bibinfo  {journal} {Journal of Applied Physics}\ }\textbf {\bibinfo {volume} {67}},\ \bibinfo {pages} {1850} (\bibinfo {year} {1990})}\BibitemShut {NoStop}%
\bibitem [{\citenamefont {Janzen}\ \emph {et~al.}(2001)\citenamefont {Janzen}, \citenamefont {Dumkow},\ and\ \citenamefont {Horn-von Hoegen}}]{janzen2001}%
  \BibitemOpen
  \bibfield  {author} {\bibinfo {author} {\bibfnamefont {A.}~\bibnamefont {Janzen}}, \bibinfo {author} {\bibfnamefont {I.}~\bibnamefont {Dumkow}},\ and\ \bibinfo {author} {\bibfnamefont {M.}~\bibnamefont {Horn-von Hoegen}},\ }\bibfield  {title} {\bibinfo {title} {Thermal activation of dislocation array formation},\ }\href {https://doi.org/10.1063/1.1408599} {\bibfield  {journal} {\bibinfo  {journal} {Applied Physics Letters}\ }\textbf {\bibinfo {volume} {79}},\ \bibinfo {pages} {2387} (\bibinfo {year} {2001})}\BibitemShut {NoStop}%
\bibitem [{\citenamefont {LeGoues}\ \emph {et~al.}(1994)\citenamefont {LeGoues}, \citenamefont {Reuter}, \citenamefont {Tersoff}, \citenamefont {Hammar},\ and\ \citenamefont {Tromp}}]{LeGoues1994}%
  \BibitemOpen
  \bibfield  {author} {\bibinfo {author} {\bibfnamefont {F.~K.}\ \bibnamefont {LeGoues}}, \bibinfo {author} {\bibfnamefont {M.~C.}\ \bibnamefont {Reuter}}, \bibinfo {author} {\bibfnamefont {J.}~\bibnamefont {Tersoff}}, \bibinfo {author} {\bibfnamefont {M.}~\bibnamefont {Hammar}},\ and\ \bibinfo {author} {\bibfnamefont {R.~M.}\ \bibnamefont {Tromp}},\ }\bibfield  {title} {\bibinfo {title} {Cyclic growth of strain-relaxed islands},\ }\href {https://doi.org/10.1103/PhysRevLett.73.300} {\bibfield  {journal} {\bibinfo  {journal} {Phys. Rev. Lett.}\ }\textbf {\bibinfo {volume} {73}},\ \bibinfo {pages} {300} (\bibinfo {year} {1994})}\BibitemShut {NoStop}%
\bibitem [{\citenamefont {Rovaris}\ \emph {et~al.}(2016)\citenamefont {Rovaris}, \citenamefont {Bergamaschini},\ and\ \citenamefont {Montalenti}}]{RovarisPRB2016}%
  \BibitemOpen
  \bibfield  {author} {\bibinfo {author} {\bibfnamefont {F.}~\bibnamefont {Rovaris}}, \bibinfo {author} {\bibfnamefont {R.}~\bibnamefont {Bergamaschini}},\ and\ \bibinfo {author} {\bibfnamefont {F.}~\bibnamefont {Montalenti}},\ }\bibfield  {title} {\bibinfo {title} {Modeling the competition between elastic and plastic relaxation in semiconductor heteroepitaxy: From cyclic growth to flat films},\ }\href {https://doi.org/10.1103/PhysRevB.94.205304} {\bibfield  {journal} {\bibinfo  {journal} {Physical Review B}\ }\textbf {\bibinfo {volume} {94}},\ \bibinfo {pages} {205304} (\bibinfo {year} {2016})}\BibitemShut {NoStop}%
\bibitem [{\citenamefont {Bauer}\ and\ \citenamefont {Schäffler}(2006)}]{bauer2006}%
  \BibitemOpen
  \bibfield  {author} {\bibinfo {author} {\bibfnamefont {G.}~\bibnamefont {Bauer}}\ and\ \bibinfo {author} {\bibfnamefont {F.}~\bibnamefont {Schäffler}},\ }\bibfield  {title} {\bibinfo {title} {Self-assembled {S}i and {SiGe} nanostructures: New growth concepts and structural analysis},\ }\href {https://doi.org/https://doi.org/10.1002/pssa.200622405} {\bibfield  {journal} {\bibinfo  {journal} {physica status solidi (a)}\ }\textbf {\bibinfo {volume} {203}},\ \bibinfo {pages} {3496} (\bibinfo {year} {2006})}\BibitemShut {NoStop}%
\bibitem [{\citenamefont {Bulatov}\ and\ \citenamefont {Cai}(2006)}]{bulatov}%
  \BibitemOpen
  \bibfield  {author} {\bibinfo {author} {\bibfnamefont {V.}~\bibnamefont {Bulatov}}\ and\ \bibinfo {author} {\bibfnamefont {W.}~\bibnamefont {Cai}},\ }\bibfield  {title} {\bibinfo {title} {Computer simulations of dislocations}\ }(\bibinfo  {publisher} {Oxford University Press, Inc.},\ \bibinfo {year} {2006})\BibitemShut {NoStop}%
\bibitem [{\citenamefont {Myronov}(2018)}]{MYRONOV201837}%
  \BibitemOpen
  \bibfield  {author} {\bibinfo {author} {\bibfnamefont {M.}~\bibnamefont {Myronov}},\ }\bibfield  {title} {\bibinfo {title} {Molecular beam epitaxy (second edition)}\ }(\bibinfo  {publisher} {Elsevier},\ \bibinfo {year} {2018})\BibitemShut {NoStop}%
\bibitem [{\citenamefont {Bording}(2000)}]{Bording2000}%
  \BibitemOpen
  \bibfield  {author} {\bibinfo {author} {\bibfnamefont {J.~K.}\ \bibnamefont {Bording}},\ }\bibfield  {title} {\bibinfo {title} {Molecular-dynamics simulation of {G}e rapidly cooled from the molten state into the amorphous state},\ }\href {https://doi.org/10.1103/PhysRevB.62.7103} {\bibfield  {journal} {\bibinfo  {journal} {Phys. Rev. B}\ }\textbf {\bibinfo {volume} {62}},\ \bibinfo {pages} {7103} (\bibinfo {year} {2000})}\BibitemShut {NoStop}%
\bibitem [{\citenamefont {Lai}\ \emph {et~al.}(2017)\citenamefont {Lai}, \citenamefont {Zhang},\ and\ \citenamefont {Fang}}]{Lai2017}%
  \BibitemOpen
  \bibfield  {author} {\bibinfo {author} {\bibfnamefont {M.}~\bibnamefont {Lai}}, \bibinfo {author} {\bibfnamefont {X.}~\bibnamefont {Zhang}},\ and\ \bibinfo {author} {\bibfnamefont {F.}~\bibnamefont {Fang}},\ }\bibfield  {title} {\bibinfo {title} {Crystal orientation effect on the subsurface deformation of monocrystalline germanium in nanometric cutting},\ }\href {https://doi.org/10.1186/s11671-017-2047-3} {\bibfield  {journal} {\bibinfo  {journal} {Nanoscale Research Letters}\ }\textbf {\bibinfo {volume} {12}},\ \bibinfo {pages} {296} (\bibinfo {year} {2017})}\BibitemShut {NoStop}%
\end{thebibliography}%

\noindent 

\clearpage

\makeatletter

\def\maketitle{
\@author@finish
\title@column\titleblock@produce
\suppressfloats[t]}
\makeatother

\setcounter{equation}{0}
\setcounter{figure}{0}
\setcounter{table}{0}
\setcounter{page}{1}
\setcounter{section}{0}
\makeatletter
\long\def\MaketitleBox{%
  \resetTitleCounters
  \def\baselinestretch{1}%
  \begin{center}%
   \def\baselinestretch{1}%
    \Large\@title\par\vskip18pt
    \normalsize\elsauthors\par\vskip10pt
    \footnotesize\itshape\elsaddress\par\vskip36pt
    \end{center}
  }
\makeatother

\title{\hrule\vspace{1cm} SUPPLEMENTARY MATERIAL \\\vspace{1cm}\hrule\vskip0pt\vspace{1.cm} Atomistic study of dislocation formation during Ge epitaxy on Si}

\maketitle
\onecolumngrid

\renewcommand{\thefigure}{S\arabic{figure}}
\renewcommand{\thesection}{S\arabic{section}}

\section{Ge-Si intermixing}

\begin{figure}[htbp!]
\centering
\includegraphics[]{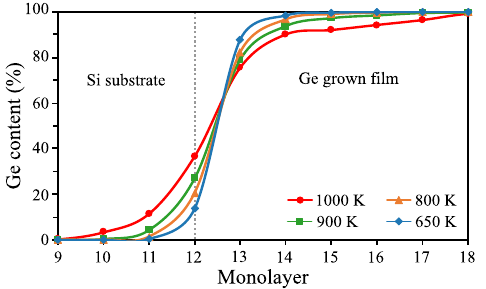}
\caption{Ge content in each ML around the original position of the Si substrate surface (vertical gray dashed line) after the deposition of 10 Ge MLs at different temperatures. For 650, 800 and 900 K, each data point is the average of the five simulated cases shown in Table 1 (dispersion in the Ge content values is negligible, so we omitted the error bars for clarity). At these temperatures, no dislocations have been formed yet, as the critical thickness is above 10 MLs (see Fig.~\ref{fig:ch5:criticalthickness}). For 1000 K, data points correspond to one particular case where a dislocation formed in order to illustrate its role in intermixing. }
\label{fig:ch5:coverage}
\end{figure}

\section{Critical thickness}

\begin{figure}[htb!]
    \centering
    \includegraphics[]{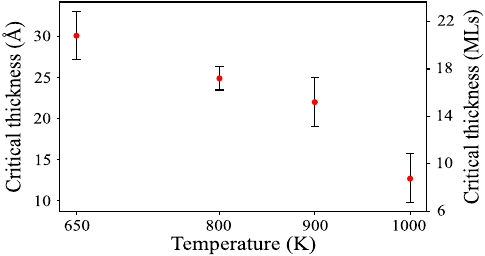}
    \caption{Average critical thickness extracted from the simulations of Table 1. For each point, error bars display the standard deviation.}
    \label{fig:ch5:criticalthickness}
\end{figure}

\newpage

\section{Stress relaxation}

\renewcommand{\thefigure}{S3.\arabic{figure}}
\setcounter{figure}{0}

The atomic stress tensor for atom $i$ with volume $V_i$ is calculated with the virial expression:

\begin{equation}
    S_{i}^{ab} = - \frac{1}{V_i} \sum_{j \neq i}(m_i v_i^a v_i^b + F_{ij}^a |r_{ij}^b|)\,,
\end{equation}

where \textit{a} and \textit{b} take the values $x$, $y$ or $z$. $m_i$ and $v_i$ are the atomic mass and velocity of atom $i$, respectively. $F_{ij}$ is the interatomic force applied on particle $i$ by particle $j$. $r_{ij}$ is the vector between particles $i$ and $j$. The atomic volume $V_i$ is computed using the \textit{Voronoi volume}. Since the Voronoi volume can not be calculated for atoms lying on the surface, we assumed for simplicity that for surface atoms $V_i$ is equal to the Voronoi volume of an atom in a perfect diamond lattice site.

The average biaxial stress per layer \textit{l}, $S_{l}^{biax}$ is computed as:
 
\begin{equation}
S^{biax}_{l} = \frac{\sum_{_{i}} \left( S_{i,l}^{xx} + S_{i,l}^{yy} \right) }{2 N^{l}}\,,
\label{eq1}
\end{equation}

where $N^l$ is the number of atoms in layer $l$ and $S_{i,l}^{xx}$ ($S_{i,l}^{yy}$) is the $xx$ ($yy$) component of the atomic stress for \textit{i}-atoms belonging to layer $l$. The normal stress component $S_{i,l}^{zz}$ is negligible as the system is free to expand along the $Z$ direction. Stress is positive for tension and negative for compression.

\begin{figure}[htb!]
    \centering
    \includegraphics[]{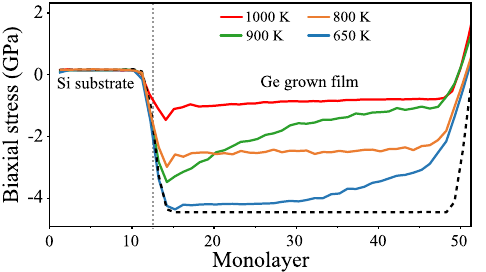}
    \caption{Distribution of the average biaxial stress per ML for the cases of Fig. 2 once deposition process is completed. The dashed black line denotes the biaxial stress of 40 Ge strained MLs coherently grown on 12 Si MLs. The vertical gray dashed line indicates the position of the original Si substrate surface. In the Si substrate region, the atoms are placed in relaxed lattice positions, thus the average biaxial stress is nearly zero. In the Si/Ge interface region, the biaxial stress shows an abrupt transition with a sharp drop to negative values (compressive stress), evidencing the lattice mismatch between Ge and Si. In the Ge grown film region above the Si/Ge interface, the biaxial stress per ML is related to the amount, type, size and arrangement of the formed dislocations in each case. From the \nth{46} ML onward, the reduction of the compressive biaxial stress and even the transition to tensile indicate the proximity of the surface. Values of the biaxial stress are not fully meaningful here due to the approximate calculation of $V_i$ for surface atoms.}
    \label{fig:ch5:stressgeneral}
\end{figure}

Figure~\ref{fig:ch5:stressgeneral} compares the biaxial stress per ML at different temperatures once the deposition process is completed. The reduction in compressive stress in the Ge grown film depends on the formed dislocation network. For instance, the dislocation network observed at 1000 K (60$^{\circ}$/90$^{\circ}$ MDs in the $Y$ direction and a Frank PD along the $X$ direction, both reaching the Si/Ge interface) releases twice the stress compared to the dislocation network formed at 800 K (two 60$^{\circ}$ MDs connected by boundary conditions that lie at the interface). The fact that dislocations reach the Si/Ge interface is crucial for stress relaxation. In the simulation at 650 K, an incipient dislocation has only propagated in the top half of the Ge grown film (up to the \nth{32} ML), and the biaxial stress underneath is very similar to the reference value. The small release of stress between \nth{15} and \nth{30} MLs is due to intermixing (negligible contribution compared to the strain released by dislocations).

\newpage

\begin{figure}[htb!]
    \centering
    \includegraphics[]{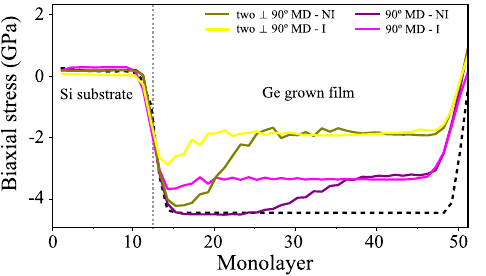}
    \caption{Distribution of the average biaxial stress per ML in samples where different 90$^{\circ}$ MDs were formed. The dashed black line denotes the biaxial stress of 40 Ge strained MLs coherently grown on 12 Si MLs. Abbreviations I and NI indicate if the dislocation line reaches the interface or not, respectively. The vertical gray dashed line shows the position of the original Si substrate surface.}
    \label{fig:ch5:90mdformation:stresscomp}
\end{figure}

To illustrate the role of 90$^{\circ}$ MDs, we have selected those simulations of Table 1 with isolated straight 90$^{\circ}$ MDs and combinations of them. Figure~\ref{fig:ch5:90mdformation:stresscomp} plots the biaxial stress per ML for the simulated cases with one or two perpendicular 90$^{\circ}$ MDs, either reaching the interface or not. The formation of a single 90$^{\circ}$ MD that reaches the interface either in $X$ or $Y$ direction (magenta line) releases approximately 1.3 GPa ($\sim$ 30\%) of the biaxial stress in each layer above the dislocation line. Two perpendicular 90$^{\circ}$ MDs that also reach the interface (yellow line) release approximately 2.7 GPa ($\sim$ 60\%) of the biaxial stress in each layer above the dislocations lines, which is double the value compared to a single 90$^{\circ}$ MD. This result is in agreement with theory since as our cell size along $X$ and $Y$ directions is $\sim$ 18 nm, the degree of relaxation in the presence of a 90$^{\circ}$ MD in each direction (i.e. two orthogonal 90$^{\circ}$ MD) at the interface should correspond to a $\sim$ 65\% relaxation of the $\sim$4\% Ge/Si lattice mismatch [11].

In both cases, either involving a single 90$^{\circ}$ MD or two perpendicular 90$^{\circ}$ MDs, if the dislocation does not reach the Si/Ge interface (dark purple or olive lines), the stress under the dislocation is high, and the transition from non-relaxed to relaxed stress regions occurs gradually. The extent of this transition is influenced by whether the dislocation line is straight or exhibits irregularities such as jogs or kinks.

\newpage

\section{Dislocation multiplication}

\renewcommand{\thefigure}{S\arabic{figure}}
\setcounter{figure}{3} 

\begin{figure}[htb!]
    \centering
    \includegraphics[width=\textwidth]{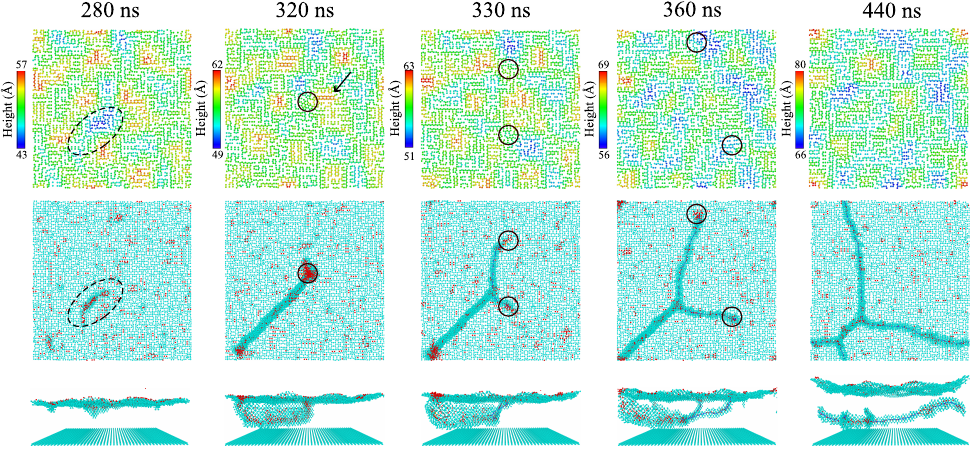}
    \caption{Sample snapshots taken at different times during one of the deposition simulations carried at 1000 K. Upper panels are top views where atoms are colored according to their height. Middle panels are also top views where atoms are colored according to their local structure: cubic diamond up to \nth{1} or \nth{2} neighbors (\protect\includegraphics[height=1.5ex]{lightblue.png}), hexagonal diamond (\protect\includegraphics[height=1.5ex]{orange.png}), and \textit{other} (\protect\includegraphics[height=1.5ex]{red.png}) (perfect cubic diamond atoms are not shown). Bottom panels are perspective views of atoms shown in middle panels. Dashed ovals indicate the valley and disordered region where the dislocation starts to form. Dark blue lines are dislocation lines as identified by OVITO. During its propagation, one of the ends of the dislocation half-loop finds a surface island (pointed by the arrow). To avoid it, the dislocation splits in two (dislocation multiplication).}
\end{figure}

\section{Post-growth annealing of a 90$^{\circ}$ MD}

\begin{figure}[htb!]
    \centering
    \includegraphics[]{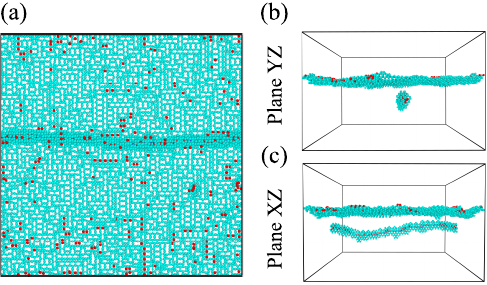}
    \caption{Snapshots taken after a 40 ns annealing at 900 K of the last sample shown in Fig. 4. (a) is a top view, while (b) and (c) are perspective views. Atoms are colored according to their local structure: cubic diamond up to \nth{1} or \nth{2} neighbors (\protect\includegraphics[height=1.5ex]{lightblue.png}), and \textit{other} (\protect\includegraphics[height=1.5ex]{red.png}) (perfect cubic diamond atoms are not shown). After annealing, the dislocation line is straight in plane XZ as kinks have been eliminated.}
\end{figure}

\end{document}